\documentclass[format=acmsmall, review=false, screen=true]{acmart}

\usepackage{booktabs} 
\usepackage[colorinlistoftodos,prependcaption,textsize=tiny]{todonotes}

\usepackage[ruled]{algorithm2e} 

\SetAlFnt{\small}
\SetAlCapFnt{\small}
\SetAlCapNameFnt{\small}
\SetAlCapHSkip{0pt}
\IncMargin{-\parindent}

\usepackage[procnames]{listings} 
\newcommand{\prettysmall}{\fontsize{6.7}{6.7}\selectfont}
\definecolor{gray98}{rgb}{0.98,0.98,0.98}
\definecolor{gray20}{rgb}{0.20,0.20,0.20}
\definecolor{gray25}{rgb}{0.25,0.25,0.25}
\definecolor{gray16}{rgb}{0.161,0.161,0.161}
\definecolor{gray60}{rgb}{0.6,0.6,0.6}
\definecolor{gray30}{rgb}{0.3,0.3,0.3}
\definecolor{bgray}{RGB}{248, 248, 248}
\definecolor{amgreen}{RGB}{77, 175, 74}
\definecolor{amblu}{RGB}{55, 126, 184}
\definecolor{amred}{RGB}{228,26,28}
\usepackage{xargs}

\usepackage{hyperref}
\usepackage{cleveref}

\newcommand{\eg}{e.g.,\ }
\newcommand{\ie}{i.e.,\ }
\newcommand{\cf}{cf.\ }
\newcommand{\sdot}{\@ifnextchar.{}{.}}

\newcommand{\gko}{\textsc{Ginkgo}\xspace}
\newcommand{\dealii}{\textsc{deal.ii}\xspace}
\newcommand{\mfem}{\textsc{MFEM}\xspace}
\newcommand{\xsdk}{\textsc{xSDK}\xspace}
\newcommand{\class}[1]{\texttt{\textit{#1}}\xspace}
\newcommand{\func}[1]{\texttt{#1}\xspace}

\newcommand{\linop}{\class{LinOp}}
\newcommand{\linopf}{\class{LinOpFactory}}

\newcommand{\exec}{\class{Executor}}
\newcommand{\stopc}{\class{Criterion}}
\newcommand{\op}{\class{Operation}}

\newcommand{\arrayc}{\class{Array}}
\newcommand{\polyobj}{\class{PolymorphicObject}}
\newcommand{\spmv}{\textsc{SpMV}\xspace}
\newcommand{\mv}{\textsc{MV}\xspace}

\newcommandx{\unsure}[2][1=]{\todo[linecolor=ACMRed,backgroundcolor=ACMRed!25,bordercolor=ACMRed,#1]{#2}}
\newcommandx{\change}[2][1=]{\todo[linecolor=ACMBlue,backgroundcolor=ACMBlue!25,bordercolor=ACMBlue,#1]{#2}}
\newcommandx{\info}[2][1=]{\todo[linecolor=ACMGreen,backgroundcolor=ACMGreen!25,bordercolor=ACMGreen,#1]{#2}}
\newcommandx{\improvement}[2][1=]{\todo[linecolor=ACMPurple,backgroundcolor=ACMPurple!25,bordercolor=ACMPurple,#1]{#2}}
\newcommandx{\thiswillnotshow}[2][1=]{\todo[disable,#1]{#2}}

\setcopyright{none}

\begin{document}
\title{\gko: A Modern Linear Operator Algebra Framework for High Performance
  Computing}

\author{Hartwig Anzt}
\orcid{0000-0003-2177-952X}
\affiliation{%
  \institution{Karlsruhe Institute of Technology}
  \city{Karlsruhe}
  \postcode{76344}
  \country{Germany}
}
\affiliation{%
  \institution{Innovative Computing Laboratory, University of
Tennessee}
  \city{Knoxville}
  \country{USA}
}
\email{hartwig.anzt@kit.edu}

\author{Terry Cojean}
\orcid{0000-0002-1560-921X}
\affiliation{%
  \institution{Karlsruhe Institute of Technology}
  \city{Karlsruhe}
  \postcode{76344}
  \country{Germany}
}
\email{terry.cojean@kit.edu}

\author{Goran Flegar}
\affiliation{%
  \institution{
    Universidad Jaime I}
  \city{12.071--Castell\'{o}n}
  \country{Spain}
}
\email{gflegar@uji.es}

\author{Fritz G\"{o}bel}
\affiliation{%
  \institution{Karlsruhe Institute of Technology}
  \city{Karlsruhe}
  \postcode{76344}
  \country{Germany}
}
\email{fritz.goebel@kit.edu}

\author{Thomas Gr\"{u}tzmacher}
\affiliation{%
  \institution{Karlsruhe Institute of Technology}
  \city{Karlsruhe}
  \postcode{76344}
  \country{Germany}
}
\email{thomas.gruetzmacher@kit.edu}

\author{Pratik Nayak}
\affiliation{%
  \institution{Karlsruhe Institute of Technology}
  \city{Karlsruhe}
  \postcode{76344}
  \country{Germany}
}
\email{pratik.nayak@kit.edu}

\author{Tobias Ribizel}
\affiliation{%
  \institution{Karlsruhe Institute of Technology}
  \city{Karlsruhe}
  \postcode{76344}
  \country{Germany}
}
\email{tobias.ribizel@kit.edu}

\author{Yuhsiang Mike Tsai}
\affiliation{%
  \institution{Karlsruhe Institute of Technology}
  \city{Karlsruhe}
  \postcode{76344}
  \country{Germany}
}
\email{yu-hsiang.tsai@kit.edu}

\author{Enrique S. Quintana-Ort\'{i}}
\affiliation{%
\institution{
    Universitat Polit\`ecnica de Val\`encia}
  \city{46.022--Valencia}
  \country{Spain}
}
\email{quintana@disca.upv.es}

\renewcommand{\shortauthors}{H. Anzt et al.}

\begin{abstract}
In this paper, we present \gko, a modern C++ math library for scientific high 
performance computing.
While classical linear algebra libraries act on matrix and vector objects, 
\gko's design principle abstracts all functionality as ``linear operators,'' 
motivating the notation of a ``linear operator algebra library.'' \gko's 
current focus is oriented towards providing sparse linear algebra functionality 
for high performance GPU architectures, but given the library design, this 
focus can be easily extended to accommodate other algorithms and hardware architectures. We 
introduce this sophisticated software architecture that separates core 
algorithms from architecture-specific backends and provide details on 
extensibility and sustainability measures. We also demonstrate \gko's usability 
by providing examples on how to use its functionality inside the MFEM and 
deal.ii finite element ecosystems. Finally, we offer a practical demonstration of 
\gko's high performance on state-of-the-art GPU architectures. 
\end{abstract}

%

\ccsdesc[500]{Mathematics of computing~Mathematical software}
\ccsdesc[499]{Computing methodologies~Massively parallel algorithms}
\ccsdesc[498]{Software and its engineering~Software creation and management}

\keywords{High Performance Computing, Healthy Software Lifecycle, Multicore and Manycore Architectures}

\maketitle

\section{Introduction}

With the rise of manycore accelerators, such as graphics processing units 
(GPUs), there is an increasing 
demand for linear algebra libraries that can efficiently transform the 
massive hardware concurrency available in a single compute node into high arithmetic performance. At the same 
time, more and more application projects adopt object-oriented 
software designs based on C++. 

In this paper, we present the result from our effort toward the design and 
development
of \gko, a next-generation, high performance sparse linear algebra library for multicore and manycore architectures. 
The library combines ecosystem extensibility with heavy, 
architecture-specific kernel optimization using the platform-native languages 
CUDA (for NVIDIA GPUs), HIP (for AMD GPUs), and OpenMP (for general-purpose multicore processors, such as those from Intel, AMD or ARM). The 
software development cycle that drives \gko ensures production-quality code by featuring unit 
testing, automated configuration and installation, 
Doxygen\footnote{\url{http://www.doxygen.nl/}} code 
documentation, 
as well as a continuous integration and continuous benchmarking framework. 
\gko is an open source effort licensed under the BSD 
3-clause.\footnote{\url{https://opensource.org/licenses/BSD-3-Clause}}

\begin{figure}[t]
  \begin{center}
    \includegraphics[width=0.98\textwidth]{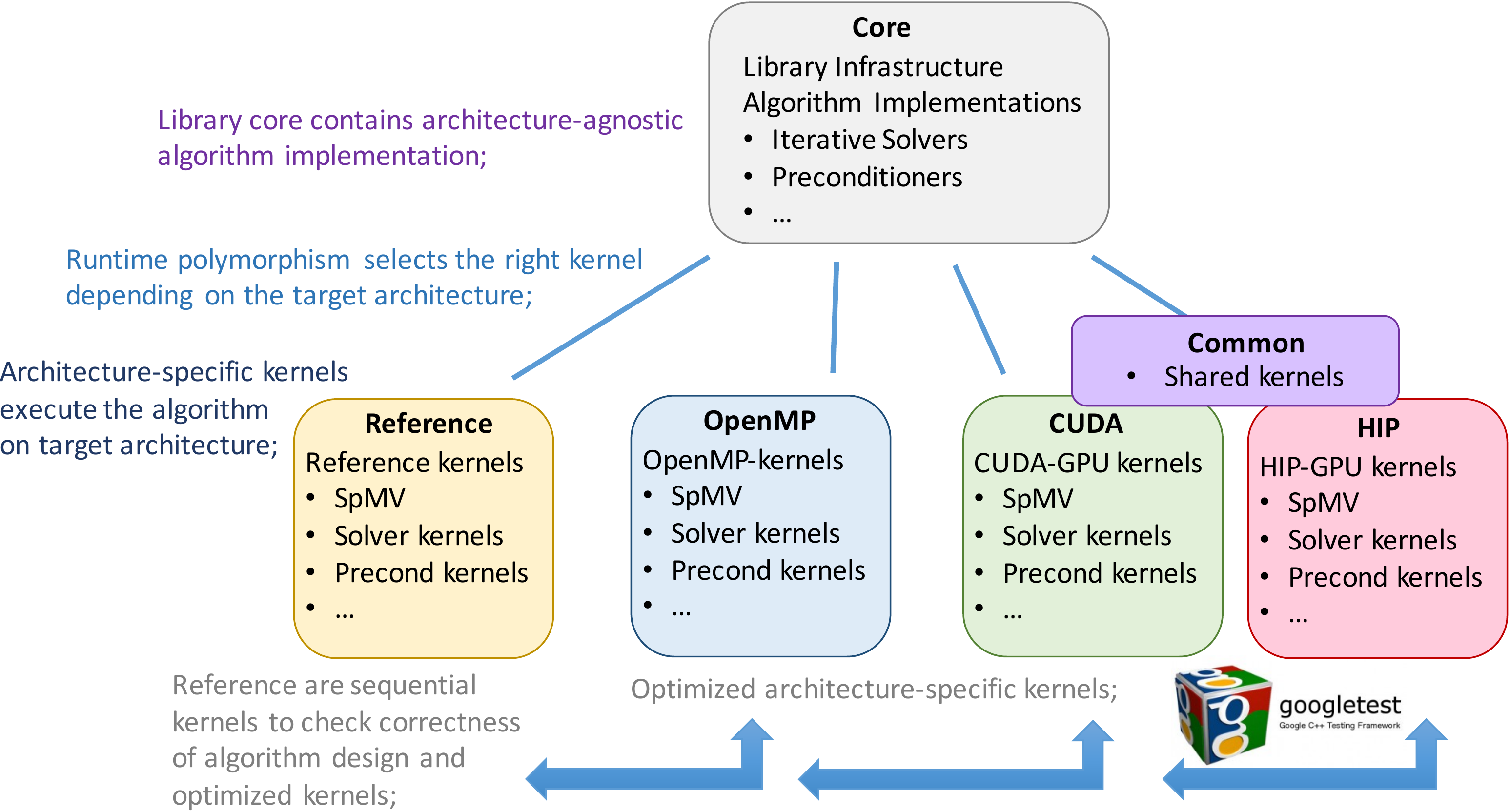}
  \end{center}
  \caption{\gko library architecture separating the core containing the 
  algorithms from architecture-specific backends.}
  \label{fig:ginkgooverview}
\end{figure}

The object-oriented \gko library is constructed around two principal design concepts.
The first principle, aiming at future technology readiness, is to consequently 
separate the numerical algorithms from the hardware-specific kernel 
implementation to ensure correctness (via comparison with sequential 
reference kernels), performance portability (by applying hardware-specific 
kernel optimizations), and extensibility (via kernel backends for other 
hardware architectures), see Figure~\ref{fig:ginkgooverview}. The second design 
principle, aiming at 
user-friendliness, is the convention to express functionality in terms of linear 
operators: every solver, preconditioner, factorization, matrix-vector product, and matrix 
reordering is expressed as a linear operator (or composition thereof).

The rest of the paper is organized as follows.
In \Cref{sec:design}, we leverage a simple use case to motivate the design
choices underlying \gko, and elaborate on the concept of linear operators, memory management,
hardware-specific kernel optimization, and event logging. \Cref{sec:solvers}
provides additional details on \gko's current solvers, realizations for the 
sparse matrix-vector product (\spmv) kernel, and preconditioner
capabilities. \Cref{sec:extensibility} elaborates on how the design allows
for easy extension, so that users can contribute new algorithmic technology or additional
hardware backends. As many applications are in desperate need for high
performance sparse linear algebra technology, \Cref{sec:external_libraries}
showcases the usage of \gko as a backend library in scientific applications, and also
reviews \gko's integration into the extreme-scale Software Development Kit (xSDK). 
In~\Cref{sec:sustainability} we describe how \gko's design and development cycle promotes sustainable software development;
and in~\Cref{sec:evaluation}, we offer representative performance results indicating \gko's competitiveness for sparse linear algebra on high-end GPU architectures.
We conclude in~\Cref{sec:conclusion} with a summary of the paper and the 
potential of the library design becoming a role model for future developments.

\section{An Overview of Ginkgo's design}
\label{sec:design}

\begin{figure}[t]
  \begin{center}
    \includegraphics[width=0.9\textwidth]{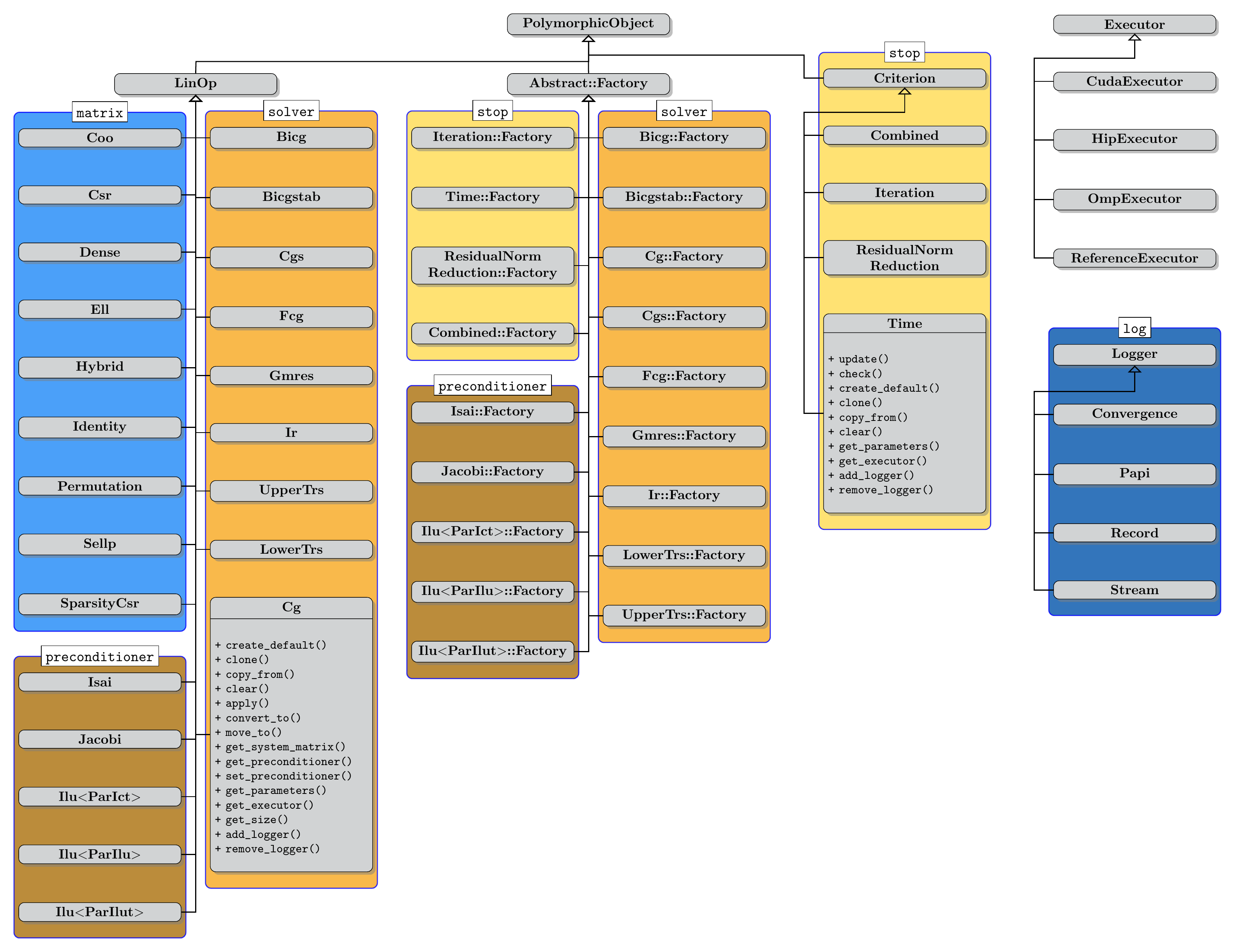}
  \end{center}
  \caption{\gko's class hierarchy showcasing the main namespaces (colored
    boxes) and classes (gray boxes) for \gko.}
  \label{fig:gko-hierarchy}
\end{figure}

\Cref{fig:gko-hierarchy} displays \gko's rich class hierarchy together with
its main namespaces and classes. To better understand the role of each object,
this section introduces \gko's interface using a minimal, concrete example as
a starting point, and gradually presenting more advanced abstractions that demonstrate
\gko's high composability and extensibility.
These abstractions include:
\begin{itemize}
  \item the \linop and \linopf classes which are used to implement and compose
    linear algebra operations, 
  \item the \exec classes that allow transparent algorithm execution on multiple
    devices; and
  \item other utilities such as the \stopc classes, which control the iteration
    process, as well as the memory passing decorators that allow fine-grained
    control of how memory objects are passed between different components of 
    the library
    and the application. 
\end{itemize}

\subsection{\gko usage example}
\label{sec:intro/example}

\begin{figure}[tbh]
  \begin{center}
    \includegraphics[width=0.9\textwidth]{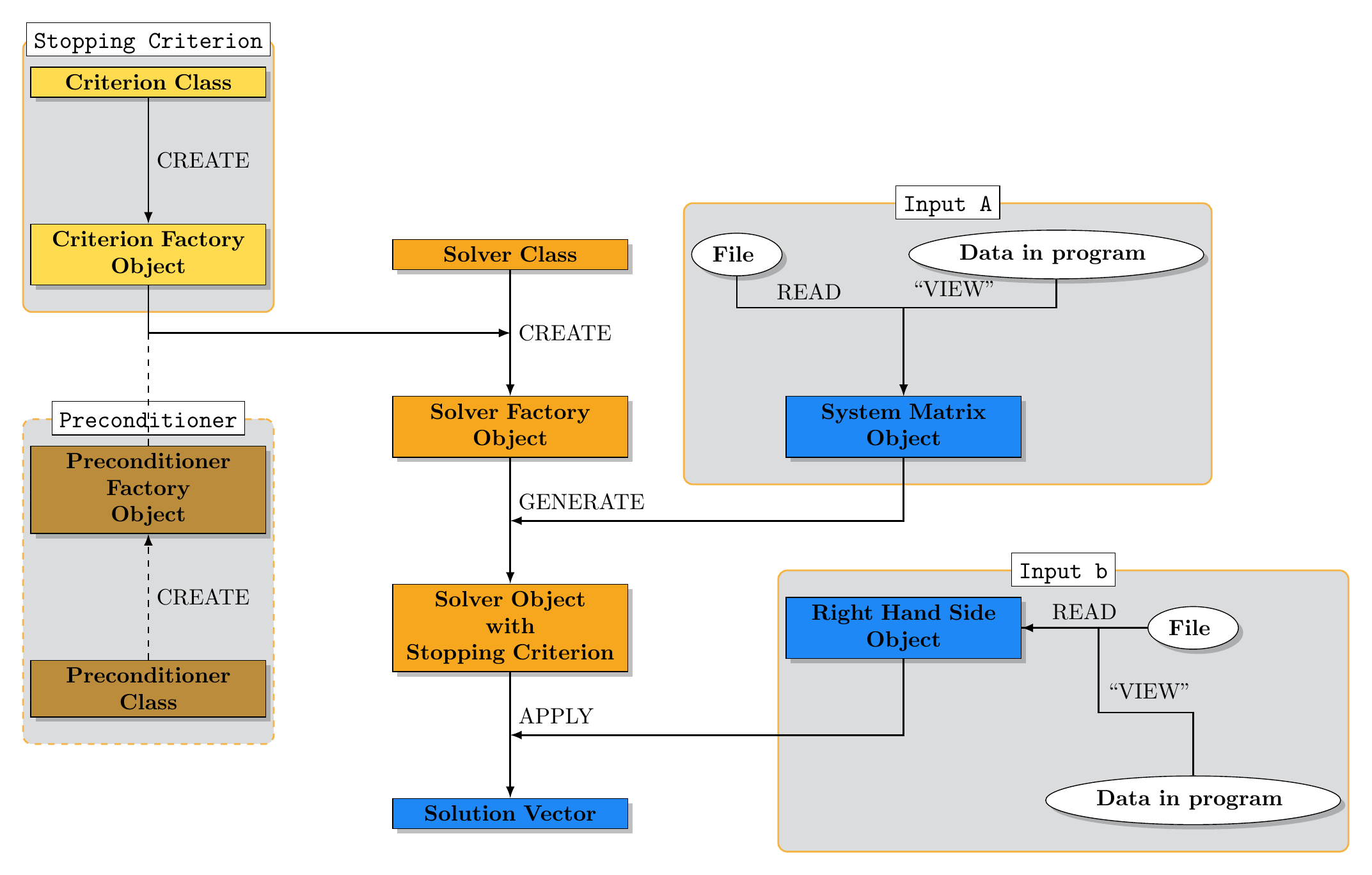}
  \end{center}
  \caption{Flowchart providing an alternative view of the code example shown in
    \Cref{lst:gko_code}. All object interactions are represented by
    arrows. The colors correspond to the type of the objects following the color convention 
    in \Cref{fig:gko-hierarchy}.}
  \label{fig:gko-flowchart}
\end{figure}

\Cref{fig:gko-flowchart} illustrates the specific 
flowchart \gko uses to solve a linear system, highlighting the interactions 
between \gko's classes. In the program code for this example given in 
\Cref{lst:gko_code}, the system matrix \texttt{A}, the right-hand side 
\texttt{b}, and the initial solution guess \texttt{x}, are
initially read from the standard input 
using \gko's `read' utility 
(lines~9--11). 
Next, the program creates a factory for a CG Krylov solver preconditioned with 
a block-Jacobi scheme
(lines~13--15).
The solver is configured to stop either
after 20 iterations or having improved the original residual by 15 orders of
magnitude
(lines~16--19). (Stopping criteria are further discussed in 
\Cref{sec:intro/criteria}.)
The system matrix is bound to the iterative solver, which is used to solve the 
system with the right-hand side 
and initial guess. 
The initial guess is overwritten with the computed solution (line~23). 
Solvers (and more generally \linop and \linopf) are discussed in detail in \Cref{sec:intro/linop}. 
Finally, the solution is printed to the standard output (line~25).

\begin{center}
\begin{minipage}{\linewidth}
\begin{lstlisting}[caption={A minimal example that uses \gko to solve a linear
    system. The system matrix, right-hand side, and the initial solution guess are 
    read from the standard input. The system is solved on an 
    NVIDIA-enabled GPU using the
    CG method enhanced with a block-Jacobi preconditioner. Two stopping criteria
    are combined to limit the maximum number of iterations and set the desired relative
    error. The solution is written to the standard output.}, label=lst:gko_code]
#include <iostream>
#include <ginkgo/ginkgo.hpp>

int main()
{
    // Instantiate a CUDA executor
    auto cuda = gko::CudaExecutor::create(0, gko::OmpExecutor::create());
    // Read data
    auto A = gko::read<gko::matrix::Csr<>>(std::cin, cuda);
    auto b = gko::read<gko::matrix::Dense<>>(std::cin, cuda);
    auto x = gko::read<gko::matrix::Dense<>>(std::cin, cuda);
    // Create the solver factory
    auto solver_factory =
        gko::solver::Cg<>::build()
            .with_preconditioner(gko::preconditioner::Jacobi<>::build().on(cuda))
            .with_criteria(
                gko::stop::Iteration::build().with_max_iters(20u).on(cuda),
                gko::stop::ResidualNormReduction<>::build()
                    .with_reduction_factor(1e-15)
                    .on(cuda))
            .on(cuda);
    // Create the solver from the factory and solve the system
    solver_factory->generate(gko::give(A))->apply(gko::lend(b), gko::lend(x));
    // Write result
    write(std::cout, gko::lend(x));
}
\end{lstlisting}
\end{minipage}
\end{center}

\gko supports execution on GPU and CPU architectures using different
backends (currently, CUDA, HIP, and OpenMP). To accommodate this, 
when creating an object, the user passes an instance of an \exec in order to 
specify where the data for that object should be stored and the operations on that data should be performed.
The particular example in \Cref{lst:gko_code} creates a CUDA \exec (line~7) that employs the first GPU device (the one
returned by \texttt{cudaGetDevice(0)}). Since CUDA GPU accelerators are
controlled by the CPU, an OpenMP \exec is needed to orchestrate the execution 
on the GPU.
(\Cref{sec:intro/executor} describes the executors model in more detail.)

\gko avoids expensive memory movement and copies. At the same time, sharing 
data between different modules in the code might cause unexpected results
(\eg one module changes a matrix used by a solver in a different module, which
causes that solver to tackle the wrong system). \gko resolves the dilemma by
allowing both shared and exclusive (unique) ownership of the objects. This 
comes at the price
of some verbosity in argument passing: in most cases, plain
arguments cannot be passed directly, but have to be wrapped in special
``decorator'' functions that specify in which ``mode'' they are passed (shared,
copied, etc.).

The minimal example in \Cref{lst:gko_code} already utilizes two of the decorator functions,
\texttt{gko::give} and \texttt{gko::lend}, both in line~23. The first one, \texttt{gko::give(A)},
causes the caller to yield the ownership of matrix \texttt{A} to the solver,
leaving the caller's version of \texttt{A} in a valid, but undefined state (\eg
accessing any of its methods is not defined, but the object can still be
de-allocated or assigned to). The second decorator, appearing twice, in \texttt{gko::lend(x)} and \texttt{gko::lend(b)},
``lends'' objects \texttt{x} and \texttt{b} to the solver by temporarily
passing ownership to it until the control flow returns from \texttt{apply} back
to the caller. This is a special ownership mode that is only used when the
callee does not need permanent ownership of the object.  Different ownership
modes, as well as their relation to \texttt{std::move} are discussed in
\Cref{sec:intro/memory-management}.

\subsection{\linop and \linopf}
\label{sec:intro/linop}

\subsubsection{Motivation}
\label{sec:intro/linop/motivation}

\gko exposes an application programming interface
(API) that allows to easily combine different components for the iterative
solution of linear systems: solvers, matrix formats, preconditioners, etc.
The API enables running distinct iterative solvers and enhancing the solvers 
with different types of preconditioners.
A preconditioner can be a matrix or even another solver. Furthermore, the system matrix does not need to be stored explicitly
in memory, but can be available only as a function that is applied to a
vector to compute a matrix-vector product (matrix-free). The objective of providing 
a clean and easy-to-use interface mandates that all these special cases are 
uniformly realized in the API.

The central observation that guides \gko's design is that the operations and
interactions between the solver, the system matrix, and the preconditioner can
be represented as the application of \textit{linear operators}:

\begin{enumerate}
  \item The major operation that an iterative solver performs on the system matrix
    \texttt{A} is the multiplication with a vector (realized as a Matrix-Vector
    product, or \mv). This operation can be viewed as the application of the induced
    linear operator $L_A : z \mapsto Az$. In some cases, multiplication with the
    transpose is also needed, which is yet another application of a linear
    operator $L_{A^T} : z \mapsto A^Tz$.
  \item The solver itself solves a system $Ax = b$, which is the application of
    the linear operator $S_A : b \mapsto A^{-1}b (= x)$. Here, the term
    ``solver'' is not used to denote a function $f$ that takes $A$ and $b$ as
    inputs and produces $x$, but instead a function with the system matrix $A$
    already fixed (that is, $S_A = f(A\,, \cdot)$).
  \item The application of the preconditioner $M$, as in $v = M^{-1}u$, can
    be viewed as the application of the linear operator $P_M : u \mapsto
    M^{-1}u (=v)$.
\end{enumerate}

There are several remarks that have to be made regarding the observations above.
First, in the context of numerical computations, with finite precision
arithmetic, the term ``linear operator''
should be understood loosely. In fact, none of the previous categories
strictly satisfy the linearity definition of the linear operator:
$L(\alpha x + \beta y) = \alpha L(x) + \beta L(y)$, where $\alpha,\beta$ are scalars and $x,y$ denote vectors. Instead, they are just
approximations of the linear operators that satisfy the formula
$L(\alpha x + \beta y) = \alpha L(x) + \beta L(y) + E$, where the error term
$E = E(L, \alpha, \beta, x, y)$ is the result of one or more of the following
effects:
\begin{enumerate}
  \item rounding errors introduced by storing non-representable values in
    floating-point format;
  \item rounding errors introduced by finite-precision floating-point 
  arithmetic;
  \item instability and inaccuracy of the method used to apply the linear 
  operator to a vector; and
  \item inexact operator application, e.g. only few iterations of an iterative 
  linear solver.
\end{enumerate}

The data layout and the implementation of any linear operator is 
internal to that operator, and the interface does not expose implementation 
details. For example, a direct solver could store its matrix data in factored 
form, as
two triangular factors (\eg $A=LU$) and implement its application as two
triangular solves (with $L$ and $U$). In contrast, an iterative solver could
just store the original system matrix, and the entire implementation of the method could
be a part of the linear operator application. Nonetheless, both operators can
still expose the same public interface.

\subsubsection{\linop}
\label{sec:intro/linop/linop}

In coherence with the observations in \Cref{sec:intro/linop/motivation}, the
central abstraction in \gko's design is the abstract class (interface) \linop,
which represents 
the mathematical concept of a linear operator. All concrete linear operators (solvers, matrix formats,
preconditioners) are instances of \linop. Furthermore, this generic operator $L$ exposes a
pure virtual method \texttt{apply(b,~x)} that is overridden by a
concrete linear operator with an implementation that computes the result
$x = L(b)$ with conformal dimensions for $L$, $x$ and $b$, where vectors are interpreted as dense matrices of dimension $n\times 1$.
This design enables that a single
interface can be leveraged to compute an {\mv} with different matrix
formats, the application of distinct types of preconditioners, the solution of
linear systems using various solvers, or even the application of a user-defined
linear operator.

Using the \linop abstraction, an iterative solver can be implemented
via references to other {\linop}s that represent the system matrix and the
preconditioner. The solver does not have to be aware of the type of the matrix
or the preconditioner --- it is sufficient to know that they are both conformal with the
\linop interface. This means that the same implementation of the solver can be
configured to integrate various preconditioners and matrices. Furthermore, the 
linear operator abstraction can also be used to
compose ``cascaded'' solvers where the preconditioner can be replaced by
another, less accurate solver, or even to create matrix-free methods by
supplying a specialized operator as the system matrix, without explicitly
storing the matrix.

\subsubsection{\linopf}
\label{sec:intro/linop/factory}

\linop exposes a uniform interface to different types of linear algebra
operations. A missing piece in the puzzle is how these {\linop}s are created in
the first place. For example, in order to solve a system with a matrix $A$ represented
by the linear operator $L_A$, an operation has to be provided which, given
the operator $L_A$, creates a solver operator $S_A$. Similarly, to create a
preconditioner $P_A$ for a matrix $A$, an operator that maps $L_A$ to $P_A$ is
needed. These are both examples of higher-order (non-linear) functions that
map linear operators to other linear operators (in this case $\Sigma : L_A
\mapsto S_A$  and $\Phi : L_A \mapsto P_A$). \gko provides an abstract class
\linopf that represents mappings such as $\Sigma$ and $\Phi$. Concretely, the class \linopf provides
an abstract method \texttt{generate(\linop)} which, given a linear operator
from the domain of the mapping, returns the corresponding \linop from its
input.

The linear operators constructed by using operator factories are usually solvers
and preconditioners. For example, in order to construct a BiCGSTAB solver operator that
solves a problem with the system matrix $A$, represented by the operator $L_A$, one
would first create a BiCGSTAB factory (which implements the \linopf interface
and represents the operator $S$); and then call \texttt{generate} on $S$,
passing the operator $L_A$ as input, to obtain a BiCGSTAB operator $S_A$, with the system
matrix, $A$.

Some factories are designed to be combined with other factories. For instance, to create an
iterative refinement solver, which uses CG preconditioned with Jacobi as the 
inner solver, one would create an
iterative refinement factory $S$, and as the inner solver factory, pass a CG
factory constructed with a Jacobi factory as the preconditioner factory.
Then, when calling the \texttt{generate} method on $S$ with the system matrix
represented by a linear operator $L_A$, this linear operator is propagated to the
CG and Jacobi factories, to create CG and Jacobi operators with the system
matrix $A$.

Instead of using \linopf, an alternative (and more obvious) approach would
have been to just use the constructor of \linop to provide all the ``component''
linear operators. However, this alternative presents the drawback that the ``type''
of the operator cannot be decoupled from its data. To illustrate this, consider
the scenario of a solver $S$ which tackles a linear system using the LU
factorization; and then invokes two triangular solvers on the resulting $L$ and
$U$ factors. There are multiple algorithms for the solution of the
triangular systems, which in \gko are represented by different linear operators.
Thus, the operators to use should somehow be passed as input parameters to the
solver $S$. The problem is that they cannot be constructed outside of $S$, since
their factors are not known at that point. \linopf provides an elegant
solution to this problem, since instead of a \linop, the solver $S$ can be
provided with linear operator factories, which are then used to construct the
triangular solver operators once the factors $L$ and $U$ are known.

\subsubsection{Re-visiting the example}
\label{sec:intro/linop/example}

After the previous elaboration on \linop and \linopf, it is timely to re-visit the example in
\Cref{lst:gko_code}. The objects
\texttt{A}, \texttt{b} and \texttt{x} in lines~9--11 are \linop objects that store their data as ``matrices'' in CSR (compressed sparse row~\cite{saad_iter_methods})
and dense matrix formats, respectively. Calling the method \texttt{apply} on
these objects 
has the effect of calculating the matrix-vector product using that data.  The
\texttt{solver\_factory} object (defined in lines~13--21), is actually a compound
\linopf used to create a solver with the CG
method. In this particular case, the CG solver is preconditioned with a block-Jacobi method
(specified by providing a block-Jacobi factory as the preconditioner factory to
the CG factory).

All the work actually occurs in line~23. First, the CG factory \texttt{solver\_factory}
is used to generate a linear operator object representing the CG solver by calling the
\texttt{generate} method. Since \texttt{solver\_factory} has a block-Jacobi factory set
as the preconditioner factory, the \texttt{solver\_factory}'s \texttt{generate} method
invokes \texttt{generate} on the block-Jacobi factory; and the system
matrix \texttt{A} is passed as input argument, which has the effect of generating a block-Jacobi
preconditioner operator for that matrix. Then, the resulting linear operator is
immediately used to solve the system by applying it on \texttt{b}. This
will have the effect of iterating the CG solver preconditioned with the
generated block-Jacobi preconditioner operator on the system matrix \texttt{A},
thus solving the system.

\subsubsection{Linear operator algebra}
\label{sec:intro/linop/algebra}

Traditional linear algebra libraries, such as BLAS~\cite{blas} and
LAPACK~\cite{laug}, use vectors and matrices as basic objects, and provide
operations such as matrix products and the solution of linear systems on these
objects as functions. In contrast, \gko achieves composability and extensibility
(\cf \Cref{sec:extensibility}) by treating linear operations as basic
objects, and providing methods to manipulate these operations in order to
express the desired complex operation. This is the principle guiding the design
of \gko, which motivates the title of this paper: while other libraries can be
characterized as ``linear algebra libraries'', \gko's algebra is  performed on
linear operators, making it a ``linear operator algebra library''.

While the current focus of \gko is on the iterative solution of sparse linear
systems, other types of operations on linear operators also fit into \gko's
concept of \linop and \linopf. For example, a matrix factorization $A = UV$
can be viewed as a linear operator factory $\Psi : L_A \mapsto F_{U, V}$, where
the linear operator $F_{U, V} : b \mapsto UVb$ stores the two factors $U$ and
$V$, and provides public methods to access the factors.

\subsection{Executors for transparent kernel execution on different devices}
\label{sec:intro/executor}
An appealing feature of \gko is the ability to run code on a variety of device architectures
transparently. In order to accommodate this functionality, \gko introduces 
the \exec class at its core. In consequence, 
the first task a user has to do when using \gko is to create an \exec.

The \exec specifies the memory location and the execution space of the linear algebra objects and
represents computational capabilities of distinct devices. 
Currently, four
executor types are provided:
\begin{itemize}
  \item \class{CudaExecutor} for CUDA-enabled GPUs;
  \item \class{HipExecutor} for HIP-enabled GPUs;
  \item \class{OmpExecutor} for OpenMP execution on multicore CPUs; and
  \item \class{ReferenceExecutor} for sequential execution on CPUs (used for
    correctness checking).
\end{itemize}
Each of these executors implements methods for allocating/deallocating memory on
the device targeted by that executor, copying data between executors, running
operations, and synchronizing all operations launched on the executor.

\Cref{lst:gko_code} illustrated the use of \exec. 
Combined with the \texttt{gko::clone(Executor, Object)} utility function, 
the \exec class makes it straight-forward to move all data and operations to a host OpenMP executor, as in
\Cref{lst:gko_matrix_copy_example}. That code creates an
\texttt{gko::OmpExecutor} object for execution on the CPU (line~1). Next, a CUDA executor
representing a GPU device with ID 0 is created (line~2); and the system matrix
data is read from a file and allocated on the 
\texttt{gko::CudaExecutor}'s device memory (line~4). Finally, the
function \func{gko::clone} creates a copy of \texttt{A} on the
\texttt{gko::OmpExecutor}, that is, in the platform's main memory (line~6). 

\begin{center}
  \begin{minipage}{\linewidth} 
    \begin{lstlisting}[caption={Copy of a matrix in CSR format from a CUDA
        device to a CPU through the \class{OmpExecutor}.},
      label={lst:gko_matrix_copy_example}]
      auto omp    = gko::OmpExecutor::create();
      auto cuda   = gko::CudaExecutor::create(0, omp);
      // As in previous example, A is allocated on a CUDA device
      auto A      = gko::read<gko::matrix::Csr<>>("data/A.mtx", cuda);
      // copy A to an OpenMP-capable device
      auto A_copy = gko::clone(omp, A);
      // All subsequent operations triggered from A_copy will use executor omp
    \end{lstlisting}
  \end{minipage}
\end{center}

\begin{figure}[t]
  \begin{center}
    \includegraphics[width=0.8\textwidth]{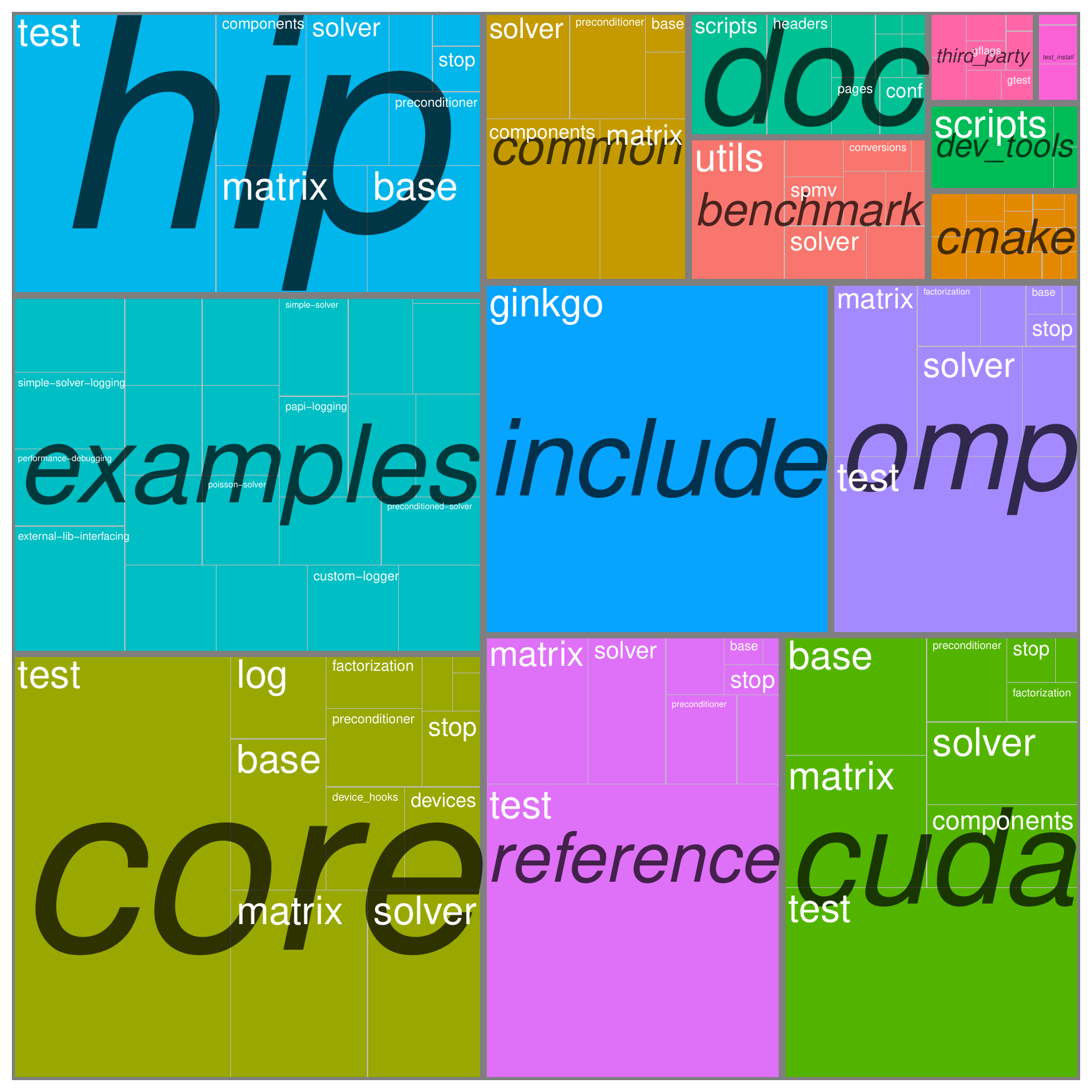}
  \end{center}
  \caption{Code distribution among different modules in \gko develop version
    1.1.1. The entire code base in this release is 8.0~MB (represented by the
    entire figure). The top level rectangles represent different top-level
    directories; these are: the \textit{core} (1.4~MB) module, examples
    (1.2~MB), the \textit{HIP} module (928~KB), the \textit{CUDA} module
    (920~KB), the \textit{reference} module (924~KB), the include directory with
    the core module's public headers (852~KB), the \textit{omp} module (612~KB),
    the \textit{common} directory which contains shared HIP and CUDA kernels
    (388~KB) and the benchmark (244~KB) and doc directories (212~KB each). The first
    rectangles in the \textit{core}, \textit{CUDA}, \textit{HIP}, \textit{omp}
    and \textit{reference} modules represent unit tests for these modules, which
    amount to 644, 396, 400, 308 and 612~KB, respectively.
  }
  \label{fig:code-distribution}
\end{figure}

In order to allow a transparent execution of operations on multiple executors,
the kernels in \gko have separate implementations for each executor type,
organized into several modules, see Figure~\ref{fig:ginkgooverview} and 
Figure~\ref{fig:code-distribution} for the code distribution, respectively. The 
\textit{core} module contains all class
definitions and non-performance critical utility functions that do not depend on
an executor. In addition, there is a module for each executor, which
contains the kernels and utilities specific for that executor. Each module is
compiled as a separate shared library, which allows to mix-and-match
modules from different sources. This paves the road for hardware vendors to
provide their own proprietary modules: they only have to optimize their module,
make it available in binary form, and users can then link it with \gko. 
We note that the similarities between HIP and CUDA allow the 
usage of \textit{common} template kernels that are identical in kernel design 
but are compiled with architecture-specific parameters to either the 
\class{HipExecutor} or the \class{CudaExecutor}. This
strategy reduces code replication and favors productivity and maintainability.

\gko contains dummy kernel implementations of all modules that throw an 
exception whenever they are called. This allows a user to deactivate certain 
modules if no hardware support is available or to reduce compilation time. In 
general, 
during the configuration step, 
\gko's automatic architecture detection activates 
all modules for which hardware support has been detected. 

The \exec design allows switching the target device where the solver in
\Cref{lst:gko_code} is executed through a one-line change that replaces
the executor used for it. In addition, if one of the arguments for the
\texttt{apply} method is not on the same executor as the operator being applied,
the library will temporarily move that argument to the correct executor before
performing the operation, and return it back once the operation is complete.
Even though this is done automatically, the user may attain higher performance 
by explicitly moving the arguments in order to avoid unnecessary copies (in the
case, for example, of repeated kernel invocation).

\subsection{Memory management}
\label{sec:intro/memory-management}

Libraries have to  specify several key memory management aspects: memory allocation, data movement and copy, and memory deallocation.
In contrast to traditional libraries such as BLAS and LAPACK, which leave memory
management to the user, \gko allocates/deallocates its memory automatically,
using the C++ ``Resource Acquisition Is Initialization'' 
(RAII\footnote{\url{https://en.cppreference.com/w/cpp/language/raii}}) 
concept combined
with the native allocation/deallocation functions of the executor (\cf
\Cref{sec:intro/executor}). Alternatively, to eliminate unnecessary
allocations and data copies, \gko's matrix formats can be configured to
use raw data already allocated and managed by the application by using 
\textit{Array views}.

A more difficult problem is to realize data movement and copies between
different entities of the application (\eg functions and other objects). 
The memory management has to not only protect against memory 
leaks or invalid memory deallocations, but also avoid unnecessary data
copies. The problem
is usually solved by specifying a well-defined owner for each object, 
responsible for deallocating the object once it is no longer needed.

\begin{figure}[t]
	\begin{center}
		\includegraphics[width=0.8\textwidth]{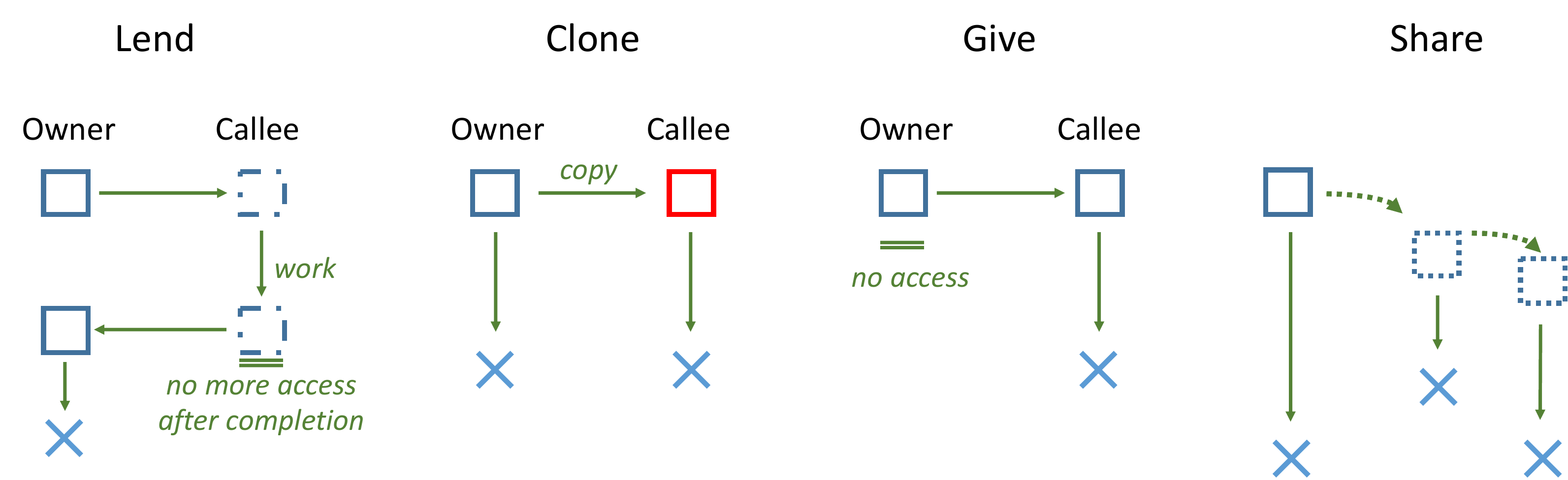}
	\end{center}
	\caption{Different ways of passing polymorphic objects as parameters in 
	\gko:\\\texttt{gko::clone}, \texttt{gko::lend}, \texttt{gko::give}, 
	\texttt{gko::share} together with the lifetime of the passed 
	object.}
	\label{fig:memory-management}
\end{figure}

For simple C++ types, this behavior is enabled via the use of
parameter qualifiers: Parameters are passed \textit{by-value} and thus copied
unless explicitly declared as references (which is when they are passed 
\textit{by-reference} without copying). The C++11 standard added \textit{move 
semantics} as a third alternative where an input parameter that is either 
explicitly (using \texttt{std::move}) or implicitly (by not having a name) 
designated a temporary value may move its internal data into the function 
without copying, leaving it in a valid but unspecified state.
However, trying to pass polymorphic objects \textit{by-value} would lead to 
object slicing~\cite{object-slicing}. In \gko, we avoid these issues with 
polymorphic types like \texttt{Executor} and \texttt{LinOp} by always passing 
and returning them as pointers. To this goal, we use the smart pointer types 
\texttt{std::unique\_ptr} and \texttt{std::shared\_ptr}, which were added in 
the C++11 standard. They provide safe resource management using RAII while 
still providing (almost) the same semantics as raw pointers.
\gko uses pointers for parameters and return types in three different contexts, 
where we say that a function parameter is used in a non-owning context if the 
object will only be used during the function call, and in an owning context if 
the object needs to be accessible even after the function call completed. 
Figure~\ref{fig:memory-management} shows the different ways to pass a 
polymorphic object as a parameter in \gko.

Functions that only need to modify a polymorphic object in a non-owning context 
take this object as a raw pointer parameter \texttt{T*}. To simplify the 
interaction with smart pointers, \gko provides the overloaded 
\texttt{gko::lend} function which returns the underlying raw pointer for both 
smart and raw pointers. This decorator function allows for a concise and 
uniform way to pass polymorphic objects to functions without ownership 
transfer. 
``Lending'' an object can be compared with normal \textit{by-reference} 
semantics for value types. When \textit{by-value} semantics are necessary, we 
can explicitly pass a copy using \texttt{gko::lend(gko::clone($\cdot$))}.

Functions that need to receive a polymorphic object in an owning context take 
this object as a \texttt{std::shared\_ptr<T>}. We can pass an object to such a 
parameter in three ways: \texttt{gko::clone} creates a copy of the current 
object to be passed to the function (\textit{by-value}), \texttt{gko::give} 
specifies that the object will not be used afterwards and can thus be moved 
into the function (\textit{move semantics}) and \texttt{gko::share} specifies 
that the ownership should be shared with the function (\textit{by-reference}). 
Note that the gko::share annotation can usually be left out, since all owning 
smart pointers in C++ already provide conversions to \textit{std::shared\_ptr}.

Functions that create new instances of a polymorphic object return a 
\texttt{std::unique\_ptr<T>}, while access to already existing objects is 
provided with \texttt{std::shared\_ptr<const T>} to allow the objects to be 
used in both owning and non-owning contexts.

The overloaded decorator functions \texttt{gko::clone}, \texttt{gko::lend}, 
\texttt{gko::give} and \texttt{gko::share} provide a uniform interface for all 
types of smart and raw pointers, while still ensuring type safety. For example, 
calling \texttt{gko::give} with a non-owning pointer will fail to compile and 
output an appropriate error message.

\subsection{Control of the iteration process}
\label{sec:intro/criteria}

Virtually all iterative methods include the concept of a ``stopping criterion'' that
evaluates whether the current approximation to the solution of the
linear systems is accurate enough. To facilitate controlling the iteration 
process, \gko provides a
collection of stopping criteria. All of them are implementations of the base
\stopc class, which specifies what type of information can be passed to the
stopping criterion. A concrete criterion provides an implementation of the
\func{check()} method that verifies if its condition has been met and,
therefore, the iteration process has to be stopped.

The stopping criteria are initially generated from criterion factories (created by
the user) by passing the system matrix, right-hand side, and an initial guess. 
In addition, during the iteration process, information can be
updated when calling the \func{check()} function with the new iteration count,
residual, solution or residual norm.

Currently, three basic stopping criteria are provided in \gko:
\begin{itemize}
  \item The \class{Time} criterion, which automatically stops the iteration
    process after a certain amount of time;
  \item the \class{Iteration} criterion, which stops the iteration process once
    a certain iteration count has been reached; and
  \item the \class{ResidualNormReduction} criterion, which stops the iteration
    process once the initial relative residual norm has been reduced by the certain
    specified amount.
\end{itemize}

Additionally, \gko provides a \class{Combined} criterion, which can be used to
combine multiple criteria together through a logical--\textit{OR} operation (\texttt{|}), so that the
first subcriterion that is fulfilled stops the iteration process. This is
illustrated in lines 16--19 of \Cref{lst:gko_code}. This design implies
some stopping criteria may detain the iteration process before ``convergence'' 
is reached, in particular the \class{Time} and \class{Iteration} criteria. \gko 
provides a \class{stopping\_status} class, which can be inspected
to find out which criterion stopped the iteration process.

The \stopc class hierarchy is designed to avoid negative impact on the 
performance,
and may even improve it. For example, in case an iterative method is applied with
multiple right-hand side vectors, the \class{stopping\_status} is evaluated 
for each right-hand side individually, skipping vector updates in subsequent 
iterations for those right-hand side vectors where convergence has been achieved.

Also, all operations required to control the iteration
process can be handled inside the \stopc classes. The consequence is that, for 
most
solvers, the residual norm and related operations are computed only when using
the \class{ResidualNormReduction} criterion. Therefore, the user can combine a
solver with a simple stopping criterion to make it more lightweight or choose
a more precise but more expensive stopping criterion. In summary, \gko's design 
of
stopping criteria tries to honor the C++ philosophy of ``only paying for what
you use''.

\subsection{Event logging}
\label{sec:intro/logging}
Another utility that is provided to users in \gko is the logging of events with 
the
purpose to record information about \gko{}'s execution. This
covers many aspects of the library, such as memory allocation, executor
events, \linop events, stopping criterion events, etc. For ease of use, 
the event logging tools provide different forms of output formats, and allow the
usage of multiple loggers at once. As with the rest of \gko, this tool is
designed to be controllable, extensible, and as lightweight as possible. To 
offer support for all those capacities, the Logger infrastructure
follows the visitor and observer design patterns~\cite{gamma1994design}. 
This 
design implies a minimal
impact of logging on the logged classes and allows to accommodate any logger.

The following four loggers are currently provided in \gko:
\begin{itemize}
  \item the \class{Stream} logger, which logs the events to a
    stream (\eg file, screen, etc.);
  \item the \class{Record} logger, which stores the events in a
    structure which has a history of all received events that the user can
    retrieve at any moment;
  \item the \class{Convergence} logger is a simple mechanism that stores the
    relative residual norm and number of iterations of the solver on
    convergence; and
  \item the \class{PAPI SDE} logger uses the PAPI Software
    Defined Events backend~\cite{papisde} in order to enable access to 
    Ginkgo's internal
    information through the PAPI interface and tools.
\end{itemize}

Almost every class in \gko possesses multiple corresponding logging events. The
logged classes are: \exec, \op, \polyobj, \linop, \linopf and \stopc. The user has the freedom to choose whether he/she wants to log all events
or select only some of them. When an event is not selected for logging by the
user, as a result of the implementation of the logging facilities, the event is not
propagated and generates a ``no-op''.

\section{Using \gko as a library}
\label{sec:solvers}

\subsection{Solver}
\label{sec:solver}

Currently, \gko provides a list of Krylov solvers (BICG, BiCGSTAB, CG, CGS, 
FCG, GMRES) for the iterative solution of 
sparse linear systems, fixed-point methods, and direct solvers for sparse 
triangular systems such as those that appear in incomplete factorization preconditioning.
In order to generate a solver, a solver factory (of type \class{LinOpFactory})
must first be created, where 
solver control parameters, such as the stopping criterion, are set. The concrete 
solver is then generated by binding the system matrix to the solver factory. 
This allows to generate multiple solvers for distinct problems with the same 
solver settings, e.g. in time-stepping methods. 
Except for Iterative Refinement~(IR), where the internal solver can be chosen,
all iterative solvers  
have the option to attach a preconditioner of the class \class{LinOp}. 
Furthermore, all solvers implement the abstract \class{LinOp} interface, which 
not only simplifies the solver usage, but also allows to use the same notation 
for calling solvers, preconditioners, SpMV, etc.
This allows the user to compose iterative solvers by choosing another iterative solver 
as a preconditioner.

\subsection{Preconditioner}
\label{sec:preconditioner}

\gko allows any solver to be used as a preconditioner, i.e., to cascade Krylov 
solvers. Additionally, \gko features diagonal scaling preconditioners 
(block-Jacobi) as well as incomplete factorization (ILU-type) preconditioners.
As any of the other solvers, preconditioners are generated through a
\class{LinOpFactory} and implement the abstract class \class{LinOp}. 

The block-Jacobi preconditioners can switch between a ``standard'' mode and an
``adaptive precision''  
mode~\cite{anzt2019adaptive}. In the latter case, the memory precision is decoupled 
from the arithmetic precision, and the storage format for each inverted 
diagonal block is optimized to preserve the numerical properties while reducing 
the memory access cost~\cite{tomsadaptive}.

The ILU-based preconditioners can be generated by interfacing vendor libraries, 
via the ParILU algorithm~\cite{chow2015asynchronous}, or via a variant known as 
the ParILUT 
algorithm~\cite{anzt2018parilut} that dynamically adapts the sparsity 
pattern of the incomplete factorization 
to the problem characteristics~\cite{anzt2019parilut}.

For the application of an ILU-type preconditioner, \gko leverages two distinct 
solvers: one for the lower triangular matrix $L$ and 
one for the upper triangular matrix $U$. The default choices are the 
direct lower and upper triangular solvers but the user can change this to use 
iterative triangular solves.

In \Cref{lst:ilu-preconditioner-custom} we illustrate how an ILU
preconditioner can be customized in almost all aspects.
In this case, we select a CGS solver for solving the upper triangular system by 
first creating the factory in lines~18--23
and then attaching it to the preconditioner factory in lines~26--28.
Instead of relying on the internal generation of the incomplete factors, we generate them ourselves
in lines~13--15. Afterwards, we generate the ILU preconditioner in line~29.
In the end, we employ the now already generated preconditioner in line~40 with a BiCGSTAB solver.

\begin{center}
	\begin{minipage}{\linewidth}
		\begin{lstlisting}[caption={An example of creating a CG solver with ILU
        preconditioning with an iterative solver for the upper triangular factor.},
		label=lst:ilu-preconditioner-custom]
#include <iostream>
#include <ginkgo/ginkgo.hpp>

int main()
{
    // Instantiate a CUDA executor
    auto cuda = gko::CudaExecutor::create(0, gko::OmpExecutor::create());
    // Read data
    auto A = gko::read<gko::matrix::Csr<>>(std::cin, cuda);
    auto b = gko::read<gko::matrix::Dense<>>(std::cin, cuda);
    auto x = gko::read<gko::matrix::Dense<>>(std::cin, cuda);
    // Generate ILU(0) factorization
    auto ilu_factorization =
        gko::factorization::ParIlu<>::build().on(cuda)
        ->generate(A);
    // Create a custom upper solver factory
    auto upper_solver_factory =
        gko::solver::Cgs<>::build()
            .with_criteria(
                gko::stop::ResidualNormReduction<>::build()
                    .with_reduction_factor(1e-5)
                    .on(cuda))
            .on(cuda);
    // Create an ILU preconditioner factory with a CGS upper solver
    auto ilu_factory =
        gko::preconditioner::Ilu<gko::solver::LowerTrs<>, gko::solver::Cgs<>>::build()
            .with_u_solver_factory(gko::share(upper_solver_factory))
            .on(cuda);
    auto ilu_prec = ilu_factory->generate(gko::share(ilu_factorization));
    // Create the solver factory with ILU preconditioning
    auto solver_factory =
        gko::solver::Bicgstab<>::build()
            .with_criteria(
                gko::stop::ResidualNormReduction<>::build()
                    .with_reduction_factor(1e-15)
                    .on(cuda),
            .with_generated_preconditioner(gko::share(ilu_prec))
            .on(cuda);
    // Create the solver from the factory and solve the system
    solver_factory->generate(gko::give(A))->apply(gko::lend(b), gko::lend(x));
    // Write result
    write(std::cout, gko::lend(x));
}
		\end{lstlisting}
	\end{minipage}
\end{center}

\section{Using \gko as a framework}
\label{sec:extensibility}

As described in \Cref{sec:design}, \gko provides a set of generic linear
operators, including various general matrix formats, popular solvers, and
simple preconditioners. However, sparse linear algebra often includes problem-specific
knowledge. This means that, in general, a highly-optimized implementation of a
generic algorithm will still be outperformed by a carefully crafted custom algorithm
employing application-specific knowledge. To tackle this, 
\gko promotes extensibility so that users can develop their own 
implementation for specific functionality without needing to modify \gko's 
code (or recompile it).

Domain-specific extensions can be elaborated as part of the application that
uses them, or even bundled together to create an ecosystem around \gko.
Currently, this is possible for all linear operators, stopping criteria,
loggers, and corresponding factories. Adding custom data types also requires only minor
changes in a single header file and a recompilation.
The only extension that
requires more significant efforts is the addition of new architectures and 
executors.
This involves modifying a key portion of \gko as it requires the addition of
specialized implementations of all kernels for the new architecture and
executor.

In contrast to the previous section, where \gko is used as a library and the
application is built around it, this section describes how \gko can be used as
a framework in which the application inserts its own custom components to work
in harmony with \gko's built-in technology.

\subsection{Utilities supporting extensibility}
\label{sec:extensibility/utils}

\gko's facilities for memory management (\eg automatic allocation and
deallocation, or transparent copies between different executors) are designed to
simplify its use as a library. As a result, the implementation burden is
then shifted to the developers of these facilities, which are either the
developers of \gko or, in case the application using \gko needs custom
extensions, the developers of that application. To alleviate the burden
and help developers focus on their algorithms,
\gko provides basic building blocks that handle memory management and the
implementation of interfaces supported by the component being developed. 

\subsubsection{\arrayc}
\label{sec:extensibility/utils/array}

Most components in \gko have some sort of associated data, which should be 
stored together with its
executor. When copying a component, its data
should also be copied, possibly to a different executor. When the object is
destroyed, the data should be deallocated with it. Doing this manually for
every class introduces a large amount of boilerplate code, which increases the
effort of developing new components, and can lead to subtle memory leaks. 
In
addition, different devices have different APIs for memory management, so a
separate version would have to be written for each executor.

To handle these issues in a single point in code, while removing some of the burden from
the developer, \gko provides the \arrayc class. This is a container which
encapsulates fixed-sized arrays stored on a specific \exec. It supports copying
between executors and moving to another executor. In addition, it leverages the
RAII idiom\footnote{\url{https://en.cppreference.com/w/cpp/language/raii}} to 
automatically deallocate itself from the memory when it is no
longer needed.

\begin{center}
\begin{minipage}{\linewidth} 
\begin{lstlisting}[caption={Usage examples of the \arrayc class.},
label={lst:gko_array_example}]
auto omp  = gko::OmpExecutor::create();
auto cuda = gko::CudaExecutor::create(0, omp);
using arr = gko::Array<int>;

arr x(cuda, {1, 2, 3, 4});             // an array of integers on the GPU
arr cpu_x(omp, x);                     // a copy of x on the CPU
arr z(omp, 10);                        // an uninitialized array of 10 integers on the CPU

z = x;                                 // copy x from the GPU to z (on the CPU)
z.set_executor(cuda);                  // move z to the GPU

auto d[]   = {1, 2, 3, 4};
auto d_arr = arr::view(omp, 4, d);     // use existing data

auto size   = x.get_num_elems();       // get the size of x
auto x_data = x.get_data();            // get raw pointer to x's data
// Note that x_data[0] would cause a segmentation fault if called from the CPU.
// Memory used for x, cpu_x and z is automatically deallocated.
// d_arr does not try to deallocate the memory.
\end{lstlisting}
\end{minipage}
\end{center}

\Cref{lst:gko_array_example} shows some common usage examples of
arrays. Lines~5--7 display several ways of initializing the \arrayc: using an
initializer list, copying from an existing array (from a different
executor), or allocating a specified amount of uninitialized memory. The last
constructor will only allocate the memory, without calling the constructors on
individual elements, which remains the responsibility of the caller. While this
is not the usual behavior in C++, properly parallelizing the construction of
the elements in multi- and manycore systems is a non-trivial task.
Nevertheless, the elements of the arrays used in \gko are mostly \textit{trivial
types}, so there is usually no need to call the constructor in the first place.

Lines~9--10 shown in \Cref{lst:gko_array_example} illustrate how the assignment
operator can be used to copy arrays and how the executor of the array can be
changed via the \func{set\_executor} 
method. The combination of the assignment operator and the RAII idiom usually
means that classes using arrays as building blocks do not require user-defined
destructors or assignment operators, since the ones synthesized by the compiler
behave as expected (in particular, this is true for all of \gko's linear
operators, stopping criteria, and loggers).

Lines~12--13 show that \arrayc can also be used to store data in a non-owning
fashion in a \textit{view}, i.e., the data will not be de-allocated when the
\arrayc is destroyed. This feature is particularly useful when using \gko to
operate on data owned by the application or another library.

Finally, raw data stored in the \arrayc can be retrieved as shown in
Lines~15--17. The \func{get\_data} method will return a raw pointer on the
device where the array is allocated, so trying to dereference the pointer from
another device will result in a runtime error.

\subsubsection{Introduction to mixins}
\label{sec:extensibility/utils/mixins}

Most components in \gko expose a rich collection of utility functions, usually
related to conversion, object creation, and memory movement. These utilities are
usually trivial to implement, and do not differ much between components.
However, they still require that the developer implements them, which steers the
focus away from the actual algorithm development. \gko addresses this issue by
using \textit{mixins}~\cite{mixin-pattern}. Since those are neither
well-known by the community\footnote{The only mixin known to the authors is
  \class{std::enable\_shared\_from\_this} from the C++ standard library.} nor
well-supported in languages commonly used in high performance computing (\eg C,
C++, Fortran), this subsection provides a simple example where mixins are leveraged
to reduce boilerplate code. The remaining parts of \Cref{sec:extensibility}
introduce mixins provided by \gko when extending certain aspects of its ecosystem. 

As a toy example, assume there is an interface \class{Clonable}, which consists
of a single method \func{clone} exposed to create a clone of an object.  This
method is useful if the object that should be cloned is only available through
its base class (\ie the static type of the object differs from its dynamic
type). A common example where this is used is the prototype design
pattern~\cite{gof-design-patterns}. Obviously, the implementation of the
\func{clone} method should just create a new object using the copy constructor.
\Cref{lst:clonable-no-mixin} is an example implementation of such a
hierarchy consisting of three classes \class{A}, \class{B} and \class{C}.
Classes \class{A} and \class{B} directly implement \class{Clonable}, while \class{C}
indirectly implements it through \class{B}.

\begin{center}
\begin{minipage}{\linewidth}
\begin{lstlisting}[caption={An example hierarchy implementing clonable without
the use of mixins.}, label=lst:clonable-no-mixin]
struct Clonable {
  virtual ~Clonable() = default;
  virtual std::unique_ptr<Clonable> clone() const = 0;
};

struct A : Clonable {
  std::unique_ptr<Clonable> clone() const override {
    return std::unique_ptr<Clonable>(new A{*this});
  }
};

struct B : Clonable {
  std::unique_ptr<Clonable> clone() const override {
    return std::unique_ptr<Clonable>(new B{*this});
  }
};

struct C : B {
  std::unique_ptr<Clonable> clone() const override {
    return std::unique_ptr<Clonable>(new C{*this});
  }
};
\end{lstlisting}
\end{minipage}
\end{center}

The implementation of the \func{clone} method is almost identical in all
classes, so it represents a good candidate for extraction into a mixin. Mixins are
not supported directly in C++, so their implementation is handled via
inheritance, usually coupled with the Curiously Recurring Template Pattern
(CRTP)~\cite{crtp-reference}. Nevertheless, using inheritance in this context
should not be viewed as establishing a parent--child relationship between the
mixin and the class inheriting from it, but instead as the class ``including''
the generic implementations provided by the mixin.
\Cref{lst:clonable-with-mixin} shows the implementation of the same
hierarchy using the \class{EnableCloning} mixin designed to provide a generic
implementation of the \func{clone} method. The mixin relies on the knowledge of
the type of the implementer to call the appropriate constructor, which is
provided as a template parameter. The base interface implemented by the
mixin is also passed as a template parameter to allow indirect implementations,
as is the case in class \class{C}.  Once the mixin is set up, any class that
wishes to implement \class{Clonable} can just include the mixin to automatically
get a default implementation of the interface, making the class cleaner, and
removing the burden of writing boilerplate code.

\begin{center}
\begin{minipage}{\linewidth}
\begin{lstlisting}[caption={An example hierarchy implementing clonable using the
\class{EnableCloning} mixin.}, label=lst:clonable-with-mixin]
struct Clonable {
  virtual ~Clonable() = default;
  virtual std::unique_ptr<Clonable> clone() const = 0;
};

template <typename Implementer, typename Base = Clonable>
struct EnableCloning : Base {
  std::unique_ptr<Clonable> clone() const override {
    return std::unique_ptr<Clonable>(
      new Implementer{*static_cast<const Implementer*>(this)});
  }
};

struct A : EnableCloning<A> {};

struct B : EnableCloning<B> {};

struct C : EnableCloning<C, B> {};

\end{lstlisting}
\end{minipage}
\end{center}

\gko uses mixins to provide default implementations, or parts of implementations
of polymorphic objects, linear operators, various factories, as well as a few
of other utility methods. To better distinguish mixins from regular classes,
mixin names begin with the ``\class{Enable}'' prefix.

\subsection{Creating new linear operators}
\label{sec:extensibility/linop}

The matrix structure is one of the most common types of domain-specific information in
sparse linear algebra. For example, the discretization of the 1D
Poisson's differential equation with a 3-point stencil results in a tridiagonal matrix with
a value $2$ for all diagonal entries and $-1$ in the neighboring diagonals. This special
structure enables designing a matrix format which only needs to store the
two values on and below/above the diagonal. Such compact matrix formats require
far less memory than general ones, which directly translates into performance
gains in the \spmv computation.

We adopt the example of the stencil matrix to demonstrate how to implement a 
custom matrix format. The code structure is shown in
\Cref{lst:custom-matrix-format}. The actual implementations of the
OpenMP, CUDA, and reference kernels are not shown here for brevity as they do
not use any important features of \gko. A full implementation is available in
\gko's \texttt{custom-matrix-format}~example, which is included in \gko's
source distribution.

Line~1 includes the \class{EnableLinOp} mixin, which implements the entire
\class{LinOp} interface except the two \func{apply\_impl} methods. These
methods are called inside the default implementation of the \func{apply} method
to perform the actual application of the linear operator. The default
implementation of \func{apply} contains additional functionalities (executor
normalization, argument size checking, logging hooks, etc.). Thus, by using the
two-stage design with \func{apply} and \func{apply\_impl}, the implementers of
matrix formats do not have to worry about these details.
Line~2 includes the \class{EnableCreateMethod} mixin, which provides a default
implementation of the static \func{create} method. The default implementation
will forward all the arguments to the \class{StencilMatrix}' constructor,
allocate and construct the matrix using the \func{new} operator, and return a
unique pointer (\class{std::unique\_ptr}) to the constructed object.

The constructor itself is defined in lines~4--8. Its parameters are the executor
where the matrix data should be located and operations performed, the size of
the stencil, and the three coefficients of the stencil. The executor and the size
are handled by \class{EnableLinOp}, and the coefficients are stored in an
\class{Array} (defined in line~55) located on the executor used by the matrix.

Linear operators provide two variants of the \func{apply} method. The ``simple''
version performs the operation $x = Ab$ and the ``advanced'' version
for $x = \alpha Ab + \beta x$. Both of them are often used in linear algebra,
and can be expressed in terms of each other: A ``simple'' application is just
an ``advanced'' one with $\alpha = 1$ and $\beta = 0$. The ``advanced''
application can be expressed by combining $x$ and the result of ``simple''
application using the \textsc{scal} and \textsc{axpy} BLAS routines (called
\func{scale} and \func{add\_scaled} in \gko). In general, specialized 
versions result in superior performance. 
Thus, 
\gko provides both of them
separately.  However, for the sake of brevity, this example implements the
``advanced'' version in terms of the ``simple'' one (lines~14--42).

The remainder of the code (lines~15--57) contains the implementation structure
of the ``simple'' application. The input parameters contain the input vector
\class{b} and the vector \class{x} where the solution will be stored.
Each input and solution vector is represented by one column of a linear
operator.
To accommodate future extensions (\eg sparse matrix--sparse vector multiplication),
both \class{x} and \class{b} are general linear operators. However, the only
type supported by this example (and all of \gko's built-in operators) is
\class{matrix::Dense}. Downcasting these vectors to \class{matrix::Dense} is
realized in lines~15--16 using the \func{gko::as} utility, which throws an
exception if one of them is not in fact a dense matrix.

The implementation of the \func{apply} operation depends on the hardware 
architecture. The
Reference version uses a simple sequential CPU implementation; the OpenMP 
version relies on a
parallel implementation based on OpenMP; and the CUDA and HIP versions launch a 
CUDA kernel and a HIP kernel, respectively. To support all four
implementations, \gko
defines the \class{Operation} interface. An object that implements this
interface is passed to the executor's \func{run} method, which will select the
appropriate implementation depending on the executor (lines~40--41). Thus,
\class{StencilMatrix} has to define a class (called \class{stencil\_operation}
in this example, lines~18--39) which implements the \class{Operation} interface
and encapsulates the four implementations. The implementations are placed into
the four overloads of the \func{run} method: the reference version in
lines~23--25; the OpenMP version in lines~26--28; the CUDA version in
lines~29--31; and the HIP version in 
lines~32--34. References to the required data also have to be passed to
\class{stencil\_operation} so that the implementation can access it.

The new matrix format can be used instead of the CSR format in the
example in \Cref{lst:gko_code} by changing the definition of \class{A}
in line~9 as shown in line~59 of \Cref{lst:custom-matrix-format}, and
placing the definition of $A$ after the definition of \class{b}. In addition,
lines~14--15 defining the preconditioner have to be removed, since the
block-Jacobi preconditioning requires additional functionalities of the matrix
format.\footnote{\class{StencilMatrix} would have to define conversion to
  \class{matrix::Csr} for block-Jacobi preconditioning to work.}

Matrix formats are not the only linear operators that can be extended. A similar
approach can be used to define new solvers and preconditioners.

\begin{center}
\begin{minipage}{\linewidth}
\begin{lstlisting}[caption={Example implementation of a user-defined matrix
format specialized for 3-point stencil matrices.},
label=lst:custom-matrix-format]
class StencilMatrix : public gko::EnableLinOp<StencilMatrix>,
                      public gko::EnableCreateMethod<StencilMatrix> {
public:
  StencilMatrix(std::shared_ptr<const gko::Executor> exec,
                gko::size_type size = 0, double left = -1.0,
                double center = 2.0, double right = -1.0)
      : gko::EnableLinOp<StencilMatrix>(exec, gko::dim<2>{size}),
        coefficients(exec, {left, center, right}) {}

protected:
  using vec = gko::matrix::Dense<>;
  using coef_type = gko::Array<double>;

  void apply_impl(const gko::LinOp *b, gko::LinOp *x) const override {
    auto dense_b = gko::as<vec>(b);
    auto dense_x = gko::as<vec>(x);

    struct stencil_operation : gko::Operation {
      stencil_operation(const coef_type &coefficients, const vec *b,
                        vec *x)
        : coefficients{coefficients}, b{b}, x{x} {}

      void run(std::shared_ptr<const gko::ReferenceExecutor>) const override {
        // Reference kernel implementation
      }
      void run(std::shared_ptr<const gko::OmpExecutor>) const override {
        // OpenMP kernel implementation
      }
      void run(std::shared_ptr<const gko::CudaExecutor>) const override {
        // CUDA kernel implementation
      }
      void run(std::shared_ptr<const gko::HipExecutor>) const override {
        // HIP kernel implementation
      }

      const coef_type &coefficients;
      const vec *b;
      vec *x;
    };
    this->get_executor()->run(
        stencil_operation(coefficients, dense_b, dense_x));
  }

  void apply_impl(const gko::LinOp *alpha, const gko::LinOp *b,
                  const gko::LinOp *beta, gko::LinOp *x) const override {
      auto dense_b = gko::as<vec>(b);
      auto dense_x = gko::as<vec>(x);
      auto tmp_x = dense_x->clone();
      this->apply_impl(b, gko::lend(tmp_x));
      dense_x->scale(beta);
      dense_x->add_scaled(alpha, gko::lend(tmp_x));
  }

private:
  coef_type coefficients;
};

// using the matrix format:
auto A = StencilMatrix::create(exec, b->get_size()[0], -1.0, 2.0, -1.0);
\end{lstlisting}
\end{minipage}
\end{center}

\subsection{Creating new stopping criteria}
\label{sec:extensibility/stop}

Implementing new stopping criteria requires a deeper understanding of the
concept than that explained in \Cref{sec:intro/criteria}. To accommodate higher
generality, a criterion is allowed to maintain state during the execution of a
solver (\eg a criterion based on a time limit may need to record the point in time
when the solver was started). On the other hand, a linear operator may invoke a
solver multiple times, every time its \func{apply} method is called. As a
consequence, the same criterion cannot be reused for multiple runs, as the state
from the previous invocation may interfere with a subsequent run. The 
solution is to
prevent users from directly instantiating criteria. Instead, the user
instantiates a criterion factory, which is then used by the solver to create a
new criterion instance every time the solver is invoked. When creating the
criterion, the solver will pass basic information about the system being solved,
which includes the system matrix, the right-hand side, the initial guess, and
optionally the initial residual. During its execution, the solver will call the
criterion's \func{check} method to decide whether to stop the process.
This method receives a list of parameters that includes the current
iteration number, and optionally one or more of the following: the current
residual, the current residual norm, and the current solution. Based on this
information, the criterion decides, separately for each right-hand side, whether the
iteration process should be detained.

Currently, \gko includes conventional stopping criteria for iterative solvers 
based on iteration
count, execution time or residual thresholds, as well as mechanisms to combine multiple
criteria. Nevertheless, users may achieve tighter control of the iteration
process by defining their own stopping criteria.
\Cref{lst:iteration-stopping-criterion} offers a sample stopping
criterion based on the number of iterations which, even though already
available in \gko as \class{gko::stop::Iteration}, is simple enough to show in
full as part of this paper.

As mentioned in \Cref{sec:intro/criteria}, all stopping criteria,
including custom ones, should implement the \class{Criterion} interface.
In addition to the \func{check} method, the interface provides various other
utility methods which facilitate memory management. To reduce the volume of
boiler-plate code needed for new stopping criteria, \gko provides the
\class{EnablePolymorphicObject} mixin. This mixin inherits an interface supporting
memory management (in this case \class{Criterion}), and implements utility
methods related to it (line~2). For the mixin to work properly, the class being
enabled has to provide a constructor with an executor as its only parameter
(lines~21--23).

Creating a criterion factory can be simplified by using the
\func{CREATE\_FACTORY\_PARAMETERS}, \func{FACTO\-RY\_PARAMETER} and
\func{ENABLE\_CRITERION\_FACTORY} macros. The first one creates a member type
\class{parameters\_type}, which contains all of the parameters of the criterion
(lines~4--6). Each parameter is defined using the \func{FACTORY\_PARAMETER}
macro, which adds a data member of the requested name and default value, as well
as a utility method ``\func{with\_<parameter~name>}'' that can be used when
constructing the factory to set the parameter. In this case, the only parameter
is the maximum number of iterations (line~5).  Finally, the
\func{ENABLE\_CRITERION\_FACTORY} macro creates a factory member type named
\class{Factory} that uses the parameters to create the criterion. The macro also
adds a data member \class{parameters\_} which holds those parameters (line~7).
When used to instantiate a new criterion, the factory will pass itself, as well
as an instance of \class{parameters\_type}, to the constructor of the criterion.
This constructor is defined in lines~25--29.

Finally, the implementation of the criterion logic is comprised inside the \func{check}
method (lines~10--19). The current state of the solver is passed via the
\class{Updater} object. This particular criterion uses the
\class{Updater::num\_iterations} property to check whether the limit on the number of
iterations has been reached (line~13). If this is not the case, the criterion
returns \texttt{false}, 
indicating to the solver that iterative process should continue (line~14). Otherwise,
the stopping statuses of all columns are set (line~16), and the
\class{one\_changed} property is set to \texttt{true} to indicate that at least
one of the statuses changed (lines~14--17). Finally, once the iteration process
for all right-hand sides has been completed, the criterion returns \texttt{true}. The
\class{stoppingId} and the \class{setFinalized} flags are additional descriptors
that may be used to retrieve additional details about the event
that stopped the iteration process.

\begin{center}
\begin{minipage}{\linewidth}
\begin{lstlisting}[caption={An example of a stopping criterion that stops the
iteration proces once a certain iteration limit is reached.},
label=lst:iteration-stopping-criterion]
class Iteration 
  : public gko::EnablePolymorphicObject<Iteration, gko::stop::Criterion> {

  GKO_CREATE_FACTORY_PARAMETERS(parameters, Factory) {
    gko::size_type GKO_FACTORY_PARAMETER(max_iters, 0);
  };
  GKO_ENABLE_CRITERION_FACTORY(Iteration, parameters, Factory);

public:
  bool check(gko::uint8 stoppingId, bool setFinalized,
             gko::Array<stopping_status> *stop_status, bool *one_changed,
             const gko::stop::Updater &updater) override {
    if (updater.num_iterations_ < parameter_.max_iters) {
      return false;
    }
    this->set_all_statuses(stoppingId, setFinalized, stop_status);
    *one_changed = true;
    return true;
  }

  explicit Iteration(std::shared_ptr<const gko::Executor> exec)
    : gko::EnablePolymorphicObject<Iteration, gko::stop::Criterion>(
        std::move(exec)) {}

  explicit Iteration(const Factory *factory,
                     const gko::stop::CriterionArgs &args)
    : gko::EnablePolymorphicObject<Iteration, Criterion>(
        factory->get_executor()),
      parameters_{factory->get_parameters()} {}
};
\end{lstlisting}
\end{minipage}
\end{center}

\subsection{Executors and extending Ginkgo to new architectures}
\label{sec:extensibility/executors}

The executor is a central class in \gko{} that provides all important primitives
for allocating/deallocating memory on a device, transferring data to other
supported devices, and basic intra-device communication (\eg synchronization).
An executor always has a \texttt{master} executor which is a CPU-side executor
capable of allocating/deallocating space in the main memory. This concept is
convenient when considering devices such as CUDA or HIP accelerators, which
feature their own separate memory space. Although implementing a \gko{} executor
that leverages features such as unified virtual memory (UVM) is possible via
the interface, in order to attain higher performance we decided to manage all copies by
direct calls to the underlying APIs.

Support for new devices (\eg optimized versions of the library for different
architectures, new accelerators or co-processors, new programming models) in a
heterogeneous node can be added to \gko by creating new executors for those
devices. This requires 1) creating a new class
which implements the \exec interface; 2) adding kernel declarations in all
\gko{} classes with kernels for the new executor; 3) extending the internal
\texttt{gko::Operation} to execute kernel operations on the new executor; and 4)
implementing kernels for all \gko{} classes on the new architectures.
Although this is an involved process and implies modifications in multiple parts
of \gko{}, the process has been successfully executed to extend \gko{} to
support a new HIP executor. Thanks to \gko{}'s design,
most changes to \gko{}'s base classes transfer to \texttt{gko::Executor}
and its related \texttt{gko::Operation} classes. In addition, although most
matrix formats, solvers, preconditioners, and utility functions rely on kernels 
that need to be implemented to support a new execution space, a good first step 
is to declare all kernels as \texttt{GKO\_NOT\_IMPLEMENTED}. This allows to obtain a compiling 
first version featuring the new executor with kernels throwing an exception when
called. The required kernel implementations can then be progressively added without 
endangering the successful compilation of the software stack.

\section{Using \gko with external libraries}
\label{sec:external_libraries}

In this section we describe and demonstrate how to interface \gko from other libraries. Specifically, we showcase the usage 
of \gko's solver and preconditioner functionality from the \dealii\cite{dealII90} 
and
\mfem\cite{mfem2019} finite element software packages.

\subsection{Using \gko as a solver}
\label{sec:external_libraries/solver}

To use \gko as a solver in an external library, one must first adapt the data
structures of the external library to \gko's data structures. We accomplish this
by borrowing the raw data from the external library's data structures; next 
operate on this data - e.g. solve a linear system; and then return the result back to 
the application in the original data format. 

\begin{center}
\begin{minipage}{\linewidth}
  \begin{lstlisting}[caption={Usage of \gko's solver capabilities in a \dealii 
  application. The code snippet only shows the solution step and assumes that 
  the system matrix and right-hand side are available from \dealii.}, 
  label=lst:gko_dealii_code]
#include <deal.II/lac/ginkgo_solver.h>
#include <deal.II/lac/sparse_matrix.h>
#include <deal.II/lac/vector.h>
#include <deal.II/lac/vector_memory.h>

#include "../testmatrix.h"
#include "../tests.h"

#include <iostream>
#include <typeinfo>

int main()
{
    // Set solver parameters
    SolverControl control(200, 1e-6);

    const unsigned int size = 32;
    unsigned int       dim  = (size - 1) * (size - 1);

    // Setup a simple matrix
    FDMatrix        testproblem(size, size);
    SparsityPattern structure(dim, dim, 5);
    testproblem.five_point_structure(structure);
    structure.compress();
    SparseMatrix<double> A(structure);
    testproblem.five_point(A);

    Vector<double> f(dim);
    f = 1.;
    Vector<double> u(dim);
    u = 0.;

    // Instantiate a Reference executor.
    auto ref = gko::ReferenceExecutor::create();

    // Create a ginkgo preconditioner.
    auto jacobi = gko::preconditioner::Jacobi<>::build().on(ref);

    // Use ginkgo to solve the system on the gpu using the CG solver with jacobi
    // preconditioning.
    // Note that this is an additional constructor that takes in a created
    // LinOpFactory object and hence is generic.
    GinkgoWrappers::SolverCG<>     solver(control, "reference", jacobi);

    // Solves the system and copies the data back to deal.ii's solution variable.
    solver.solve(A, u, f);
}
\end{lstlisting}
\end{minipage}
\end{center}

\begin{center}
\begin{minipage}{\linewidth}
  \begin{lstlisting}[caption={Usage of \gko's solver capabilities in a \mfem 
  application.}, label=lst:gko_mfem_code]
#include "mfem.hpp"

int main() {

  .
  . // Setup the finite element space and assemble the linear
  . // and bilinear forms
  .

  OperatorPtr A;
  Vector B, X;
  a->FormLinearSystem(ess_tdof_list, x, *b, A, X, B);

  // Solve the linear system with CG + ILU from Ginkgo.

  // Instantiate a Reference executor.
  auto ref = gko::ReferenceExecutor::create();
  // Setup the preconditioner.
  auto ilu_precond =
      gko::preconditioner::Ilu<gko::solver::LowerTrs<>,
                               gko::solver::UpperTrs<>>::build()
      .on(ref);

  // Create the solver object with convergence parameters.
  GinkgoWrappers::CGSolver ginkgo_solver("reference", 1, 2000, 1e-12, 0.0,
                                         ilu_precond.release());

  // The solve method internally converts the MFEM objects to Ginkgo's
  // objects if necessary, computes the solution and returns the solution.
  ginkgo_solver.solve(&((SparseMatrix &)(*A)), X, B);

  // Get solution back to MFEM
  a->RecoverFEMSolution(X, *b, x);

  .
  . // Clean up
  .
}
\end{lstlisting}
\end{minipage}
\end{center}

\Cref{lst:gko_dealii_code,lst:gko_mfem_code} showcase the
explotation of \gko functionality in \dealii and \mfem applications.
Our main objective is to expose \gko's functionalities to the external
libraries while maintaining an uniform interface within those libraries.
The interfaces preserve the libraries' own solver interface, and take 
the executor determining the execution space as the only additional parameter. 
All data movement is handled automatically and remains transparent to the user.

\subsection{Using \gko's preconditioners }
\label{sec:external_libraries/preconditioners}

\gko provides a multitude of preconditioners on both the CPU and the GPU. An
example of such a preconditioner is the block-jacobi preconditioner. To
accomodate the use of ginkgo's preconditioners in \dealii or \mfem, an additional
constructor for each of the concrete solver classes has been provided which
takes in a \texttt{gko::LinOpFactory} as an argument. In the most general case this can
be taken to be any generic linear operator factory with an overloaded apply
implementation to serve as a preconditioner. 

\subsection{Interoperability with \xsdk}
\label{sec:external_libraries/interop}

\gko is a part of the extreme-scale Scientific Software Development Kit 
(\xsdk~\cite{xsdk}), a software stack that comprises some of the most important 
research software libraries and that is available on all US leadership 
computing facilities.
\gko is included in the \xsdk release
0.5.0~\cite{xsdk-release} which is available as a Spack metapackage. 

Within the \xsdk effort, interoperability examples with \mfem and \dealii 
showcase the \linop concept of \gko, and the use of \gko as a solver using  
partial assembly of the finite element operator within \mfem.

\section{Software sustainability efforts}
\label{sec:sustainability}

An important aspect of the \gko library is its orientation towards software
sustainability, ease of use, and openness to external contributions. Aside from \gko
being used as a framework for algorithmic research, its primary intention is to 
provide a numerical software ecosystem designed for easy adoption by the 
scientific computing community. 
This requires sophisticated design guidelines and high quality code. 
With these goals in mind, \gko follows the guidelines and policies of the 
xSDK and the 
Better Scientific Software (BSSw~\cite{bssw}) initiative. 
In order to facilitate easy adoption, \gko is open source with a
modified BSD license, which does not restrict commercial use of the software. 
The main repository is publicly available on github and only prototype 
implementations of ongoing research are kept in a private repository.
The github repository is open to external contributions through a peer-review 
concept and uses issues for bug tracking and to bolster development efforts. A 
Continuous Integration (CI) system realizes the automatic synchronization of 
repositories, and the compilation and testing of the distinct branches. The CI
is also setup to ensure quality of the library in terms of memory leaks,
threading issues, detection of bugs thanks to static code analyzers, etc.
The configuration and compilation processes are facilitated with CMake.
The testing is realized using Google Test~\cite{googletest} and comprises a
comprehensive list of unit tests ensuring the library's functionality. 
A feature spearheading sustainable high performance software development is 
\gko's Continuous Benchmarking (CB) framework. This component of \gko's ecosystem automatically runs 
performance tests on each code change; archives the performance results in a 
public git repository; and allows users to investigate the performance via an 
interactive web tool, the Ginkgo Performance 
Explorer\footnote{\url{https://ginkgo-project.github.io/gpe/}}~\cite{anzt2019towards}.
Finally, the documentation is automatically kept up-to date-with the software, 
and multiple wiki pages containing examples, tutorials, and contributor 
guidelines are available.

\section{Experimental Evaluation}
\label{sec:evaluation}

\subsection{Experimental setup}
In the performance evaluation, we consider two GPU-centric HPC nodes from different hardware vendors:
The AMD node consists of an AMD Threadripper 1920X (12 x 3.5 Ghz) CPU, 
64 GB RAM, and an AMD RadeonVII GPU. The RadeonVII GPU 
features 16 GB of main memory accessible at of 1,024 GB/s (according 
to the specifications), and has a theoretical peak of 3.4 (double precision) TFLOP/s.
The NVIDIA node is integrated into the Summit supercomputer, 
and consists of two IBM POWER9 processors and six NVIDIA Volta V100 accelerators.
The NVIDIA V100 GPUs each have a theoretical peak of 7.8 (double-precision) 
TFLOP/s 
and feature 16 GB of high-bandwidth memory (HBM2). The 
board specifications indicate a memory bandwidth of 920 GB/s for this
accelerator. We run all our experiments on a single GPU.
We note that we do not intend this to be a performance-focused paper, and 
therefore refrain from showing a comprehensive performance evaluation, but only 
show selected performance results that are representative for the common usage 
of \gko.

\subsection{The cost of runtime polymorphism}
Relying on static and dynamic polymorphism largely simplifies code maintenance 
and extendability. A common concern when using these C++ features is the 
runtime overhead induced by runtime polymorphism. Due to \gko{}'s design,
multiple runtime polymorphisms are evaluated at different levels. For example,
calling the SpMV \texttt{apply()} functionality goes through 3 polymorphism
forks: Format selection, Executor selection, and Kernel variant selection. Solvers
undergo a similar process, except that during each iteration they call multiple
kernels: an SpMV, possibly a preconditioner, etc.

To evaluate the performance impact of the multiple runtime polymorphism
branches, in \Cref{tab:solver_polymorphism_cost} 
we first measure the overhead
for all \gko{}'s solvers. The results there are obtained using a matrix of
size $1$, with an initial solution $x=0$ and the right-hand side
($b$) set to $NaN$. This allows running the full solver algorithm executing all
runtime polymorphism branches with negligible kernel execution time. We report 
results for 1.000 solver iterations averaged over 10.000 solver runs. 
\Cref{tab:solver_polymorphism_cost}
shows that the time per iteration is at most 1.5$\mu{}s$ for any of the solvers.

\begin{table}
\begin{tabular}{|r||r|r|r|r|r|}
\hline
  Solver & BiCGSTAB & CG & CGS & FCG & GMRES \\
  \hline
  \hline
  Time per iteration ($\mu{}s$) & 1.26
                                & 1.28
                                & 1.00
                                & 1.45
                                & 1.51
\\\hline
\end{tabular}
\caption{Overhead of the main Ginkgo solvers measured by averaging 10.000 solver
  runs, each doing 1.000 iterations.}
\label{tab:solver_polymorphism_cost}
\end{table}

\begin{figure}
\begin{tabular}{lr}
\includegraphics[width=0.49\columnwidth]{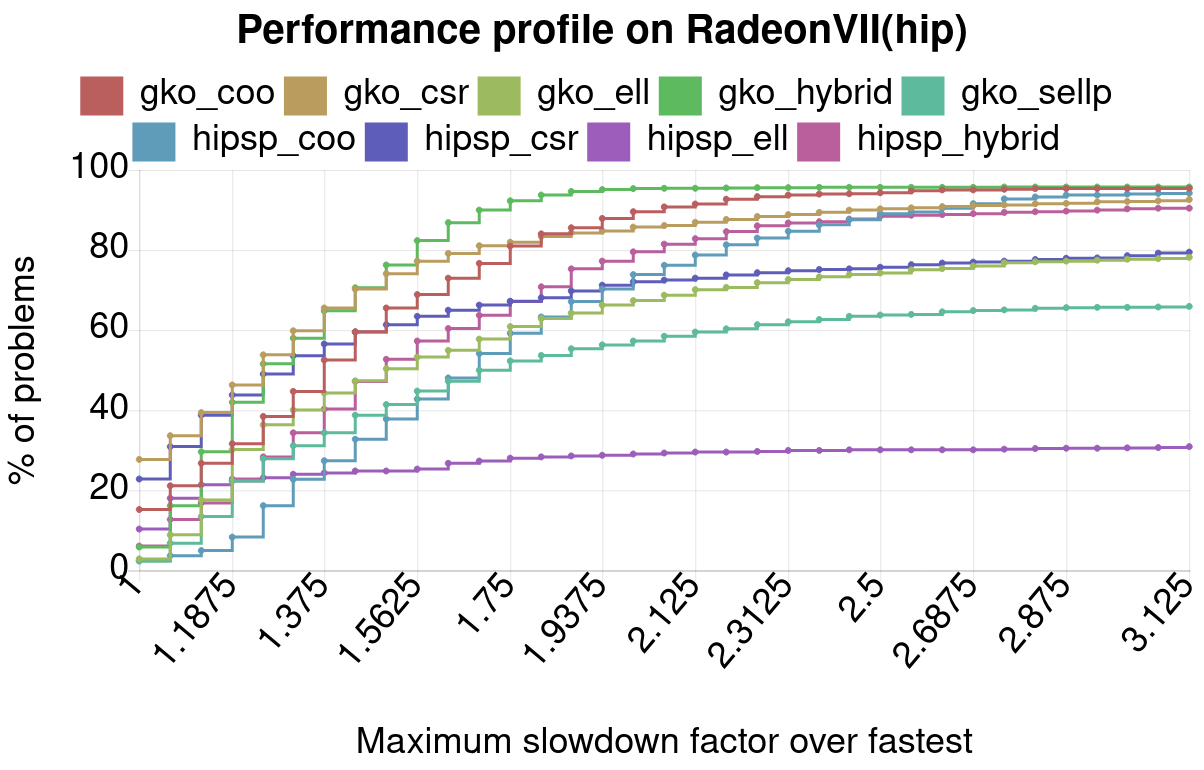} & 
\includegraphics[width=0.49\columnwidth]{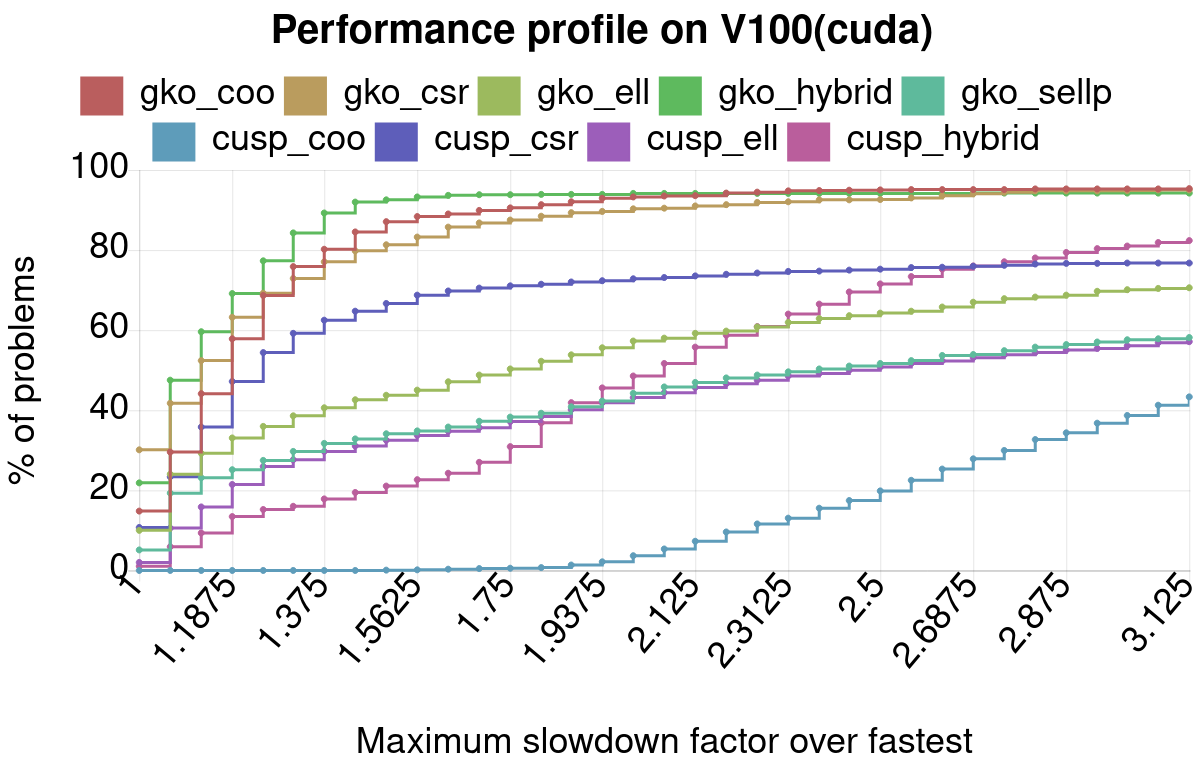}
\end{tabular}
\caption{Performance profile comparing the runtime of \gko{}'s SpMV kernels 
with the vendor libraries on the AMD RadeonVII (left) and the NVIDIA V100 
(right). The plain names represent the \gko kernels, the ``hipsp\_'' and 
``cusp\_'' labels refer to the vendor implementations in AMD's hipSPARSE and 
NVIDIA's cuSPARSE libraries, respectively.}
\label{fig:perfprofile}
\end{figure}

\subsection{SpMV kernel performance}
We next evaluate the performance of the SpMV kernel for all matrices available in 
the Suite Sparse Matrix Collection~\cite{saad_iter_methods,suitesparse} on the 
AMD RadeonVII and the NVIDIA V100 GPU~\cite{isc20}. 
For this purpose, we compare the performance
profile of the SpMV kernels available in the Ginkgo library with their
counterparts available in the NVIDIA cuSPARSE and the AMD hipSPARSE libraries.
The performance profile indicates for how many test matrices from the Suite 
Sparse Matrix Collection a specific format is the fastest (maximum slowdown 
factor 1.0), and how well a specific format generalized. I.e., for a given 
``acceptable slowdown factor,'' which percentage of the problems from the Suite 
Sparse Matrix Collection can be covered.
The performance profiles reveal that \gko{}'s 
kernels are at least competitive, 
and in many cases superior to the vendor libraries.

\begin{table}
  \begin{tabular}{l|r|r}
    Solver & Read access volume & Write access volume  \\
    \hline
    \hline
    BiCGSTAB & \begin{minipage}{5cm}$(5 \cdot n + 2 \cdot nnz) \cdot VT + 2 
    \cdot nnz \cdot IT + \lceil{}iter/2\rceil{} \cdot ((16 \cdot n + 2 \cdot 
    nnz) \cdot VT + 2 \cdot nnz \cdot IT) + \lfloor{} iter/2 \rfloor{} \cdot 
    ((13 \cdot n + 2 \cdot nnz) \cdot VT + 2 \cdot nnz \cdot IT)$ 
    \end{minipage} &\begin{minipage}{5cm} $(10 \cdot n + 6) \cdot VT + \lceil{} 
    iter/2 \rceil{} \cdot ((4 \cdot n + 2) \cdot VT)+ \lfloor{} iter/2 
    \rfloor{} \cdot ((4 \cdot n + 3) \cdot VT)$\end{minipage}\\
    \hline
    CG & \begin{minipage}{5cm}$(4 \cdot n + 2 \cdot nnz) \cdot VT  + 2 \cdot 
    nnz \cdot IT + iter \cdot ((15 \cdot n + 2 \cdot nnz) \cdot VT + 2 \cdot 
    nnz \cdot IT)$ \end{minipage} &\begin{minipage}{5cm} $(5 \cdot n + 2) \cdot 
    VT + iter \cdot ((5 \cdot n + 2) \cdot VT)$\end{minipage}\\
    \hline
    CGS & \begin{minipage}{5cm}$(5 \cdot n + 2 \cdot nnz) \cdot VT + 2 \cdot 
    nnz \cdot IT + \lceil{} iter/2 \rceil{} \cdot ((14 \cdot n + 2 \cdot nnz) 
    \cdot VT + 2 \cdot nnz \cdot IT) + \lfloor{} iter/2 \rfloor{} \cdot ((6 
    \cdot n + 2 \cdot nnz) \cdot VT + 2 \cdot nnz \cdot IT)$ \end{minipage} 
    &\begin{minipage}{5cm} $(10 \cdot n + 2) \cdot VT + \lceil{} iter/2 
    \rceil{} \cdot (6 \cdot n + 3) \cdot VT + \lfloor{} iter/2 \rfloor{} \cdot (4 \cdot 
    n \cdot VT)$\end{minipage}\\
    \hline
    FCG & \begin{minipage}{5cm}$(4 \cdot n + 2 \cdot nnz) \cdot VT + 2 \cdot 
    nnz \cdot IT + iter \cdot ((17 \cdot n + 2 \cdot nnz) \cdot VT + 2 \cdot 
    nnz \cdot IT)$ \end{minipage}& \begin{minipage}{5cm} $(6 \cdot n + 3) \cdot 
    VT + iter \cdot ((6 \cdot n + 3) \cdot VT)$\end{minipage}\\
    \hline
    GMRES & \begin{minipage}{5cm} $(11 \cdot n + 2 \cdot nnz + 5/2 \cdot r + n 
    \cdot r + r^2/2 + 1) \cdot VT + 2 \cdot nnz \cdot IT + 
    \lfloor{}iter/k\rfloor{} \cdot ((1 + 5/2 \cdot k + 10 \cdot n + 2 \cdot nnz 
    + k^2/2 + k \cdot n) \cdot VT + 2 \cdot nnz \cdot IT) + iter \cdot ((7 
    \cdot n + 5 + 2 \cdot nnz) \cdot VT + 2 \cdot nnz \cdot IT + 8) + iter_r 
    \cdot ((4 \cdot n + 4) \cdot VT)$ \end{minipage} & \begin{minipage}{5cm}$(6 
    \cdot n + r + 2 \cdot k + 3) \cdot VT + 8 + \lfloor{}iter/k\rfloor{} \cdot 
    ((k + 6 \cdot n + 2) \cdot VT + 8) + iter \cdot ((4 \cdot n + 8) \cdot VT + 
    8) + iter_r \cdot ((n + 2) \cdot VT)$ \end{minipage} \\
    \hline
\end{tabular}
\caption{Memory access volume of a full run of the distinct solver. Here $VT$
  is the value type size in bytes (e.g., for double it is 8 bytes); $IT$ is value type for the
  index type; and $iter$ is the number of iterations the solver does. In GMRES, $k$
  is the Krylov dimension (or restart iteration setting); $r = iter \% k$;
  and $iter_r = \lfloor{} iter/k \rfloor{} * (k - 1) * k / 2 + (iter\%k - 1) *
  iter\%k / 2$.}
\label{tab:solvermemcost}
\end{table}

\subsection{Ginkgo solver performance}
Prior to evaluating the performance of \gko's Krylov solvers, we point out that 
Krylov solvers operating with sparse linear systems are memory-bound 
algorithms. 
For this reason, we initially assess the bandwidth efficiency of the implementations of the different
Krylov solvers. Concretely, we select the COO matrix 
format for the SpMV kernel, and run the Krylov solvers without any preconditioner. 
In \Cref{tab:solvermemcost} we list the target Krylov solvers along with
their memory access volume (as a function of the iteration count). 
The formula for the GMRES algorithm is more involved as we implement a variant
enhanced with restart.

\begin{figure}
\begin{tabular}{lr}
\includegraphics[width=0.48\columnwidth]{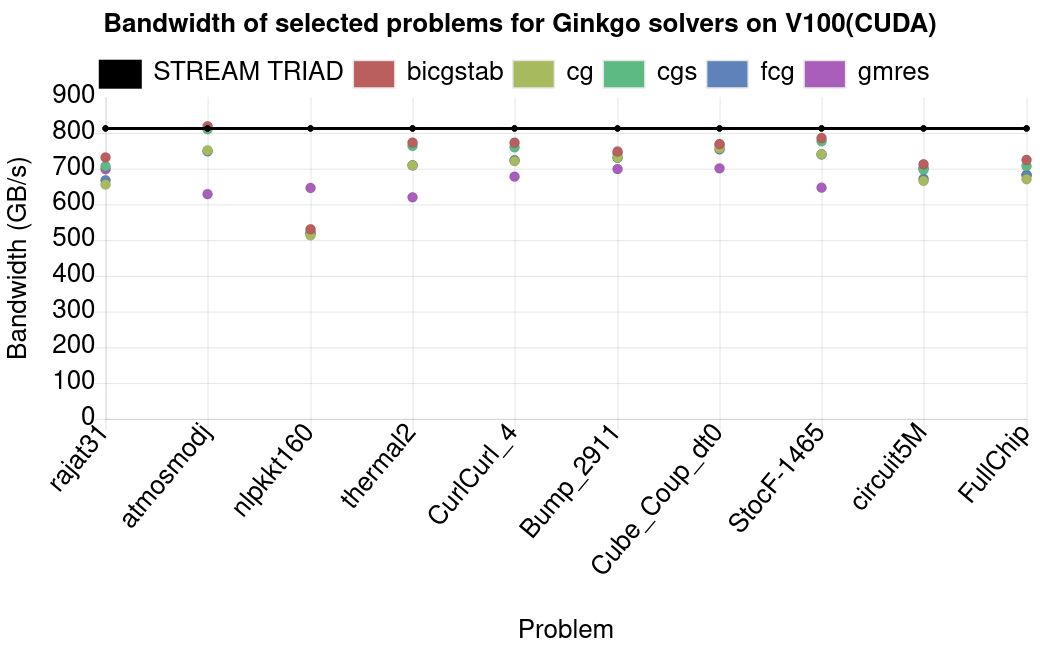} & 
\includegraphics[width=0.48\columnwidth]{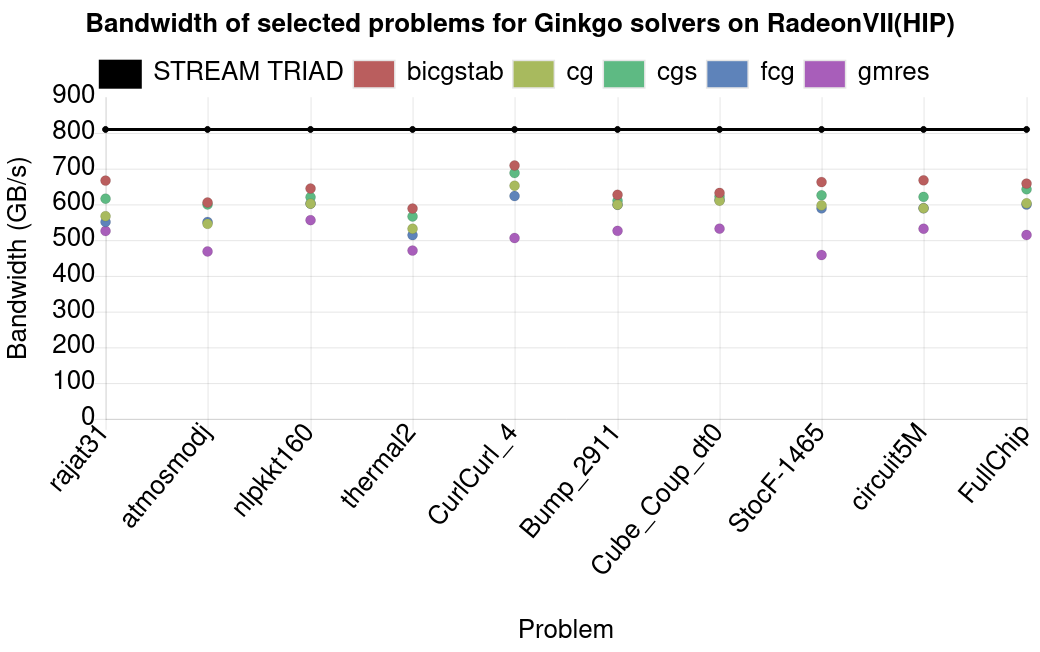}
\end{tabular}
\caption{Memory efficiency of \gko's Krylov solvers.}
\label{fig:solverbandwidth}
\end{figure}

For the experimental evaluation, we run 10,000 solver iterations on 10 
different but representative 
test matrices from the Suite Sparse collection. For GMRES, we set the restart
parameter to 100. In \Cref{fig:solverbandwidth}, we visualize the
memory bandwidth usage of the different Krylov solvers for both a V100
GPU executing CUDA code and an AMD RadeonVII GPU executing HIP code. In each
graph we indicate the experimental peak bandwidth achieved by a reference 
stream triad\footnote{$a[i]=b[i]+\alpha{}c[i]$} bandwidth 
benchmark~\cite{babelstream}.
 Both test machines reach a very similar STREAM triad bandwidth
($809.8$~GB/s on RadeonVII vs $812.6$~GB/s on V100), although the theoretical
bandwidth of the RadeonVII is higher ($1024$~GB/s on RadeonVII vs $900$~GB/s on
V100). The bandwidth performance analysis reveals that the algorithms are 
achieving bandwidth rates in the range of $500$ to $700$~GB/s on the RadeonVII 
machine and $600$ to
$800$~GB/s on the V100 machine. This means that the Ginkgo solver performance
reaches more than $70\%$ 
of the observed bandwidth on the RadeonVII machine (slightly less for GMRES), 
whereas it is more than $80\%$ on the V100 machine. To
better understand this performance discrepancy, in
\Cref{tab:stream_bandwidth_comp} we provide detailed bandwidth results on both
machines for key operations. The machines show different behaviors: for the copy
operation the RadeonVII reaches $6\%$ better performance than the V100, whereas
for the dot operation it reaches $25\%$ less performance than the V100, most
likely due to the lack of independent thread scheduling. The
relatively poor performance for global reductions on the AMD GPU may explain the
performance difference of the \gko solvers between the two machines, as global
reductions are essential components of any Krylov solver.

\begin{table}
  \begin{tabular}{l|r|r}
    Operation & V100 performance (GB/s) & RadeonVII performance (GB/s)\\
    \hline
    \hline
    Copy & 790.475 & 841.669 \\
    \hline
    Mul & 787.301 & 841.934\\
    \hline
    Add & 811.312 & 806.632\\
    \hline
    Triad & 812.617 & 809.754\\
    \hline
    Dot & 844.321 & 635.677\\
    \hline
  \end{tabular}
  \caption{Stream bandwidth results from~\cite{babelstream} on the V100 and
    RadeonVII machines for key operations.}
  \label{tab:stream_bandwidth_comp}
\end{table}

\subsection{Ginkgo preconditioner performance}
\begin{figure}
\begin{tabular}{c}
\includegraphics[width=0.6\columnwidth]{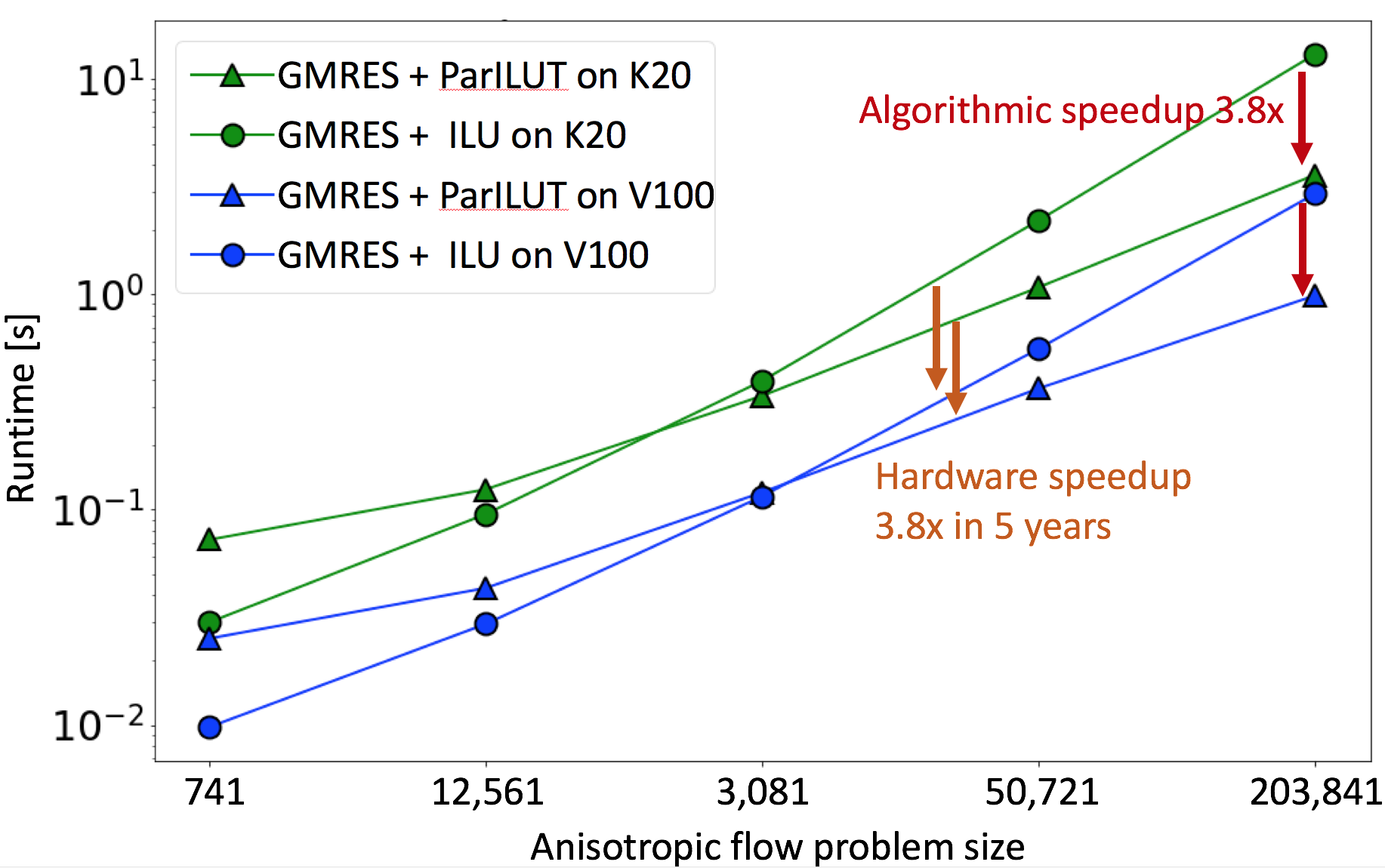}
\end{tabular}
\caption{Time-to-solution comparison between standard ILU preconditioning 
(NVIDIA's cuSPARSE) and \gko's ParILUT for solving anisotropic flow problems on 
two NVIDIA GPU generations. The GMRES solver is taken from the \gko library.}
\label{fig:parilut}
\end{figure}

\gko provides both (block-Jacobi type) preconditioners based on diagonal 
scaling and (ILU type) incomplete factorization preconditioners. \gko{}'s ILU 
preconditioner technology is spearheading the community, including ParILUT, the 
first threshold-based ILU preconditioner for GPU 
architectures~\cite{anzt2019parilut}. This preconditioner approximates the 
values of 
the preconditioner via fixed-point iterations while dynamically adapting the 
sparsity pattern to the matrix properties~\cite{anzt2018parilut}. Depending on 
the 
matrix characteristics, this preconditioner can significantly accelerate the 
solution process of linear system solves; see~\Cref{fig:parilut}.

Advanced techniques for the ILU preconditioner generation are complemented with 
fast triangular solvers, including iterative 
methods~\cite{anzt2015iterative,europar20} and the approximation of the inverse 
of 
the triangular factors via a sparse matrix (incomplete sparse approximate 
inverse preconditioning~\cite{anzt2018incomplete}).

\begin{figure}
\begin{tabular}{c}
\includegraphics[width=0.98\columnwidth]{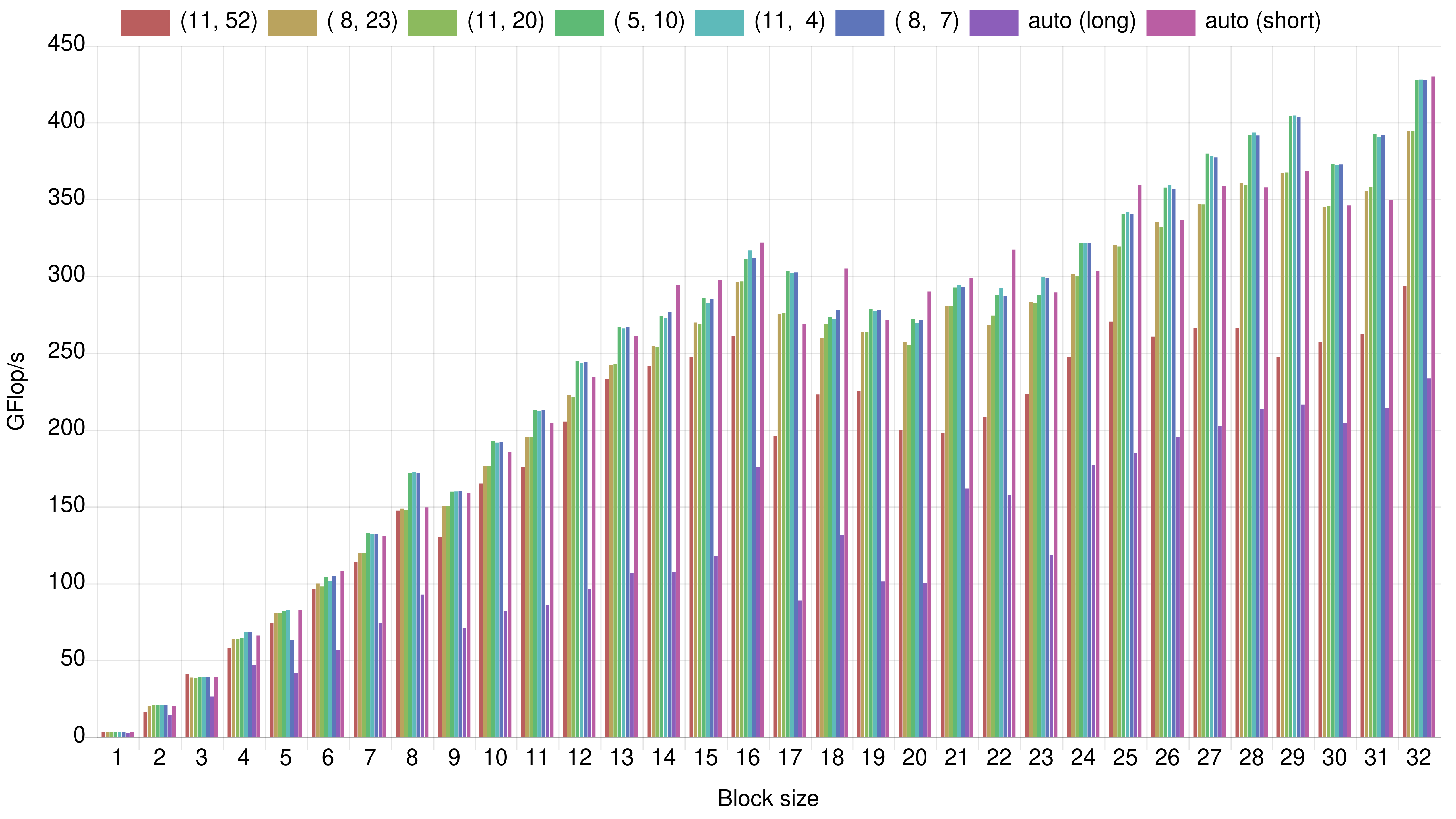}
\end{tabular}
\caption{Performance of the block-Jacobi preconditioner generation on the 
NVIDIA V100 GPU. The preconditioner generation includes the Gauss-Jordan 
elimination featuring pivoting, the condition number calculation and exponent 
range analysis, the storage format optimization, the format conversion, and the 
preconditioner storage in GPU main memory. The Different bars for each size 
represent the distinct memory precision scenarios and the scenarios where the 
performance data includes the automatic precision detection based on the block 
condition number and exponent range.}
\label{fig:bgje}
\end{figure}

The block-Jacobi preconditioner available in \gko outperforms its competitors 
by automatically adapting the memory precision to the numerical requirements, 
therewith reducing the memory access time of the memory-bound preconditioner 
application~\cite{anzt2019adaptive,tomsadaptive}. The inversion of the diagonal 
block is realized via a heavily-tuned batched variable size Gauss-Jordan 
elimination~\cite{anzt2019variable}; see~\Cref{fig:bgje}.

\section{Conclusions and perspectives}
\label{sec:conclusion}
\gko is a modern C++-based sparse linear 
algebra library for GPU-centric HPC architectures with many appealing features including
the Linear Operator abstraction, which fosters easy adoption of the library.
Some other aspects that we have elaborated on include the execution control, 
memory management via smart pointers, 
software quality measures, and library extensibility. We have also provided example 
codes for software integration into the deal.ii and MFEM finite element 
ecosystems, and demonstrated the high performance of Ginkgo on high end GPU 
architectures. We believe that the design of the library and the 
sustainability measures that are taken as part of the development process have 
the potential to become role models for the 
future efforts on scientific software packages.

\section*{Acknowledgments}
This work was supported by the ``Impuls und Vernetzungsfond of the 
Helmholtz Association" under grant VH-NG-1241.
G. Flegar and E. S. Quintana-Ort\'{\i} were supported
by project TIN2017-82972-R of the MINECO and FEDER
and the H2020 EU FETHPC Project 732631 ``OPRECOMP''.
This research was also supported by the Exascale Computing Project 
(17-SC-20-SC), a collaborative effort of the U.S. Department of Energy Office 
of Science and the National Nuclear Security Administration.
The authors want to acknowledge the access to the  Summit supercomputer at the 
Oak Ridge National Lab (ORNL).

\bibliographystyle{ACM-Reference-Format}
\bibliography{bibliography}

\end{document}